\documentclass[twocolumn]{aastex631}

\usepackage[T1]{fontenc}
\usepackage{graphicx}    % Including figure files
\usepackage{amsmath}     % Advanced maths commands
\usepackage{amssymb}     % Extra maths symbols

\usepackage{morefloats}
\def\h2{H$\rm _2$}

\def\maL{$\langle t\rangle_L$}
\def\maM{$\langle t\rangle_M$}
\def\mzL{$\langle Z\rangle_L$}
\def\mzM{$\langle Z\rangle_M$}

\def\str{{\sc starlight}}

\newcommand\footnoteref[1]{\protected@xdef\@thefnmark{\ref{#1}}\@footnotemark}

\shorttitle{The stellar content of BAH galaxies}
\shortauthors{Riffel et al.}

\begin{document}

%%%%%%%%%%%%%%%%%%% TITLE PAGE %%%%%%%%%%%%%%%%%%%

\title{Blowing star formation away in AGN hosts (BAH) - V: The Feeding-Feedback Cycle in local AGNs as revealed by their stellar populations}

\correspondingauthor{Rog\'erio Riffel}
\email{riffel@ufrgs.br}

\author[0000-0002-1321-1320]{Rog\'erio Riffel}
\affiliation{Departamento de Astronomia, Instituto de F\'isica, Universidade Federal do Rio Grande do Sul, CP 15051, 91501-970, Porto Alegre, RS, Brazil}

\author[0000-0003-0483-3723]{Rogemar A. Riffel}
\affiliation{Departamento de F\'isica, CCNE, Universidade Federal de Santa Maria, Av. Roraima 1000, 97105-900, Santa Maria, RS, Brazil}

\author[0009-0005-0583-5773]{Gabriel L. Souza-Oliveira}
\affiliation{Departamento de F\'isica, CCNE, Universidade Federal de Santa Maria, Av. Roraima 1000, 97105-900, Santa Maria, RS, Brazil}

\author[0000-0001-6100-6869]{Nadia L. Zakamska}
\affiliation{Department of Physics \& Astronomy, Johns Hopkins University, Bloomberg Center, 3400 N. Charles St, Baltimore, MD 21218, USA}

\author[0000-0003-1772-0023]{Thaisa Storchi-Bergmann}
\affiliation{Departamento de Astronomia, Instituto de F\'isica, Universidade Federal do Rio Grande do Sul, CP 15051, 91501-970, Porto Alegre, RS, Brazil}

\author[0000-0002-6570-9446]{Marina Bianchin}
\affiliation{Instituto de Astrof\'isica de Canarias, Calle V{\'\i}a L{\'a}ctea s/n, E-38205 La Laguna, Tenerife, Spain}
\affiliation{Departamento de Astrof\'isica, Universidad de La Laguna, E-38206, La Laguna, Tenerife, Spain}

\author[0009-0004-5666-0881]{Lais Nery Marinho}
\affiliation{Departamento de Astronomia, Instituto de F\'isica, Universidade Federal do Rio Grande do Sul, CP 15051, 91501-970, Porto Alegre, RS, Brazil}

\author[0009-0006-7814-2334]{Sandra Jaison}
\affiliation{Departamento de Astronomia, Instituto de F\'isica, Universidade Federal do Rio Grande do Sul, CP 15051, 91501-970, Porto Alegre, RS, Brazil}

\author[0009-0008-5705-2908]{Suresh Parekh}
\affiliation{Departamento de Astronomia, Instituto de F\'isica, Universidade Federal do Rio Grande do Sul, CP 15051, 91501-970, Porto Alegre, RS, Brazil}

\author[0009-0008-2184-1403]{Mait\^e S. Z. de Mellos}
\affiliation{Departamento de F\'isica, CCNE, Universidade Federal de Santa Maria, Av. Roraima 1000, 97105-900, Santa Maria, RS, Brazil}

\author[0000-0003-3667-9716]{Jos\'e Henrique Costa-Souza}
\affiliation{Departamento de F\'isica, CCNE, Universidade Federal de Santa Maria, Av. Roraima 1000, 97105-900, Santa Maria, RS, Brazil}

\author[0000-0001-7207-4584]{Rafael S. de Souza}
\affiliation{Departamento de Astronomia, Instituto de F\'isica, Universidade Federal do Rio Grande do Sul, CP 15051, 91501-970, Porto Alegre, RS, Brazil}
\affiliation{Centre for Astrophysics Research, University of Hertfordshire, College Lane, Hatfield, AL10~9AB, UK}
\affiliation{Department of Physics \& Astronomy, University of North Carolina at Chapel Hill, NC 27599-3255, USA}

% Abstract of the paper
\begin{abstract}
We present a spatially resolved analysis of the stellar populations in the inner kiloparsec of NGC~3884, 3C~293, and CGCG~012-070. Using near-infrared spectroscopy, we reconstruct their star formation histories (SFHs) by comparing the M13, XSL, and FSPS stellar population synthesis models. The stellar light is dominated by intermediate-age to old populations ($t\gtrsim$1\,Gyr) with super-solar metallicities ($Z \gtrsim 1\,Z_{\odot}$). All models clearly indicate recent star formation (rejuvenation) in these AGN hosts, with young to intermediate-age populations contributing significantly in the nuclear regions. The SFHs from M13 and XSL broadly agree in showing coexisting old and young components, whereas FSPS favours a larger fraction of very young ($t<$ 50\,Myr) stars. Moreover, XSL- and FSPS-based SFHs are generally more irregular and “bumpy,” while M13 yields smoother, more continuous SFHs. In NGC~3884 and 3C~293, stars with $0.2 < t \leq 0.7$~Gyr form a ring-like structure around the nucleus. The nuclear spectra further require non-stellar components: a featureless power-law continuum ($FC$) and hot dust emission ($HD$). In 3C~293, the $FC$ component appears in two spatially separated regions, possibly indicating a dual active galactic nucleus, though a heavily reddened starburst origin for the secondary component cannot be excluded. Nearly all fits show a central drop in stellar metallicity, consistent with inflow of metal-poor gas that fuels recent accretion and AGN activity. Radial profiles show that $HD$ and $FC$ contributions decrease with radius, while younger stellar populations become more prominent outward. Together, these results support a feeding–feedback scenario in which gas inflows trigger circumnuclear star formation and, via stellar mass loss, help sustain ongoing AGN activity.

.
\end{abstract}

% Keywords (AASTeX uses a unified command format)

%%%%%%%%%%%%%%%%%%%%%%%%%%%%%%%%%%%%%%%%%%%%%%%%%%

%%%%%%%%%%%%%%%%% BODY OF PAPER %%%%%%%%%%%%%%%%%%

\section{Introduction}

One of the primary challenges in modern extragalactic astrophysics is to understand the mechanisms that regulate the growth of the supermassive black holes (SMBHs). An unresolved question concerns the transport of gas from the kiloparsec scales of the host galaxy to the circumnuclear region at parsec scales, a process crucial for fueling the central supermassive black hole and triggering its Active Galactic Nucleus \citep[AGN; e.g.][]{Heckman+14, Storchi-Bergmann+19, Silva-Lima+22}. Although external triggers such as galaxy mergers can supply large fuel reservoirs, internal processes are essential to sustain the long lived AGN activity \citep[]{Davies+07,Heckman+14, Riffel+21}. Among the internal mechanisms, the recycled stellar gas, the material returned to the interstellar medium from an evolving stellar population, has emerged as a key potential fuel source \citep[e.g.][]{Davies+07, Riffel+09, Riffel+22, Riffel+24, Choi+24}. This local gas source overcomes the theoretical constraint of the large angular momentum barrier \citep[e.g.][]{Silva-Lima+22, Riffel+22, Choi+24}, offering an acceptable mechanism to sustain the low to moderate accretion rates typical of nearby AGN \citep[e.g.][]{Heckman+14, Davies+15}.  This connection between the stellar life cycle and the fueling of the central SMBH offers an excellent framework for interpreting the observed relationship between SMBH growth and the formation of their host galaxies \citep[]{Heckman+14, Madau+14}.

The connection between central black hole mass and host galaxy bulge properties, first established through the fundamental correlations between SMBH mass and stellar velocity dispersion, establishes the coevolutionary paradigm of AGN accretion and star formation (SF) processes \citep[e.g.][]{Magorrian+98, Ferrarese+00, Gebhardt+00, Hilbe+17}. Remarkably, the mass accretion requirements for sustaining AGN activity are relatively modest. A typical Seyfert galaxy operating at a bolometric luminosity of $L_{\rm bol} \cong 10^{45} \, \mathrm{erg \, s^{-1}}$ can be sustained for a Myr through the accretion of only $ \approx 2 \times 10^{5} \, M_\odot$ of material \citep[][and references therein]{Rosario+18}, making stellar mass loss a viable long term fuel source. This perspective underscores why understanding the detailed mechanics of gas transport from kiloparsec to parsec scales remains a central challenge in the field.

The observed coexistence of nuclear SF with AGN activity provides convincing proof for a complex relationship between the formation of new stars and gas accretion into the SMBH \citep[e.g.][]{Storchi-Bergmann+01, CidFernandes+04, Riffel+07, Heckman+14, Madau+14, Gatto+25}. Theoretical models and numerical simulations imply that gas inflows into the nucleus can simultaneously trigger both star formation and SMBH accretion \citep[e.g.][]{DiMatteo+05, Hopkins+10, Zubovas+13, Zubovas+17, Weinberger+23, Mercedes-Feliz+23}. A key element in this coevolution is AGN feedback, which manifests through several mechanisms, including jets, winds, radiation, and powerful outflows. On one hand, feedback suppresses star formation, which can be further categorized as either preventive, where it limits the availability of cold gas reservoirs, or ejective, where preexisting cold molecular gas is removed \citep[e.g.][]{Granato+04, Fabian+12, King+15, Trussler+20}. On the other hand, feedback can be positive, where the passage of the outflow compresses the interstellar medium, acting as a catalyst to trigger local SF \citep[e.g.][]{Rees+89, Hopkins+12, Zubovas+17, Maiolino+17}. Recent advanced cosmological hydrodynamical simulations, such as the Feedback In Realistic Environment (FIRE) project, reveal that both positive and negative feedback effects can operate simultaneously in different parts of the galaxy, with preventive mechanisms typically dominating over ejective processes in regulating galaxy-wide star formation \citep[e.g.][]{Hopkins+14, Hopkins+18, Mercedes-Feliz+23, Alexander+25}.

Observationally, the high fraction of young to intermediate age stellar populations within the central regions of AGN is firmly established \citep[e.g.][]{Terlevich+90, Storchi-Bergmann+01, CidFernandes+04, Davies+07, Riffel+07, Davies+09, Riffel+09, Martins+10, Riffel+11c, Storchi-Bergmann+12, Esquej+14, Riffel+16b, Ruschel-Dutra+17, Hennig+18, Mallmann+18, Burtscher+21, Riffel+21, Dahmer-Hahn+22, Bessiere+22, Riffel+22, Riffel+23}. However, a major outstanding puzzle is the relative timing between the burst of SF and the activation of AGN. The inherently brief AGN phase ($\sim 0.1-10$ Myr; \citealt{Novak+11, Schawinski+15}) compared to typical nuclear starburst ages ($\sim 1-100$ Myr; \citep{Hickox+14}) creates substantial challenges for chronological comparisons and leaves temporal relationships prone to large uncertainties. Although numerous investigations have identified apparent connections between younger stellar ages and higher AGN luminosity, suggesting synchronous fueling processes \citep[]{CidFernandes+04, Riffel+07, Riffel+09, Mallmann+18, Riffel+22, Riffel+23}, other detailed analyses present conflicting evidence. Notably, the comprehensive optical study of the Type 2 Low Luminosity AGN and Matched Analogues (LLAMA) sample by \citet{Burtscher+21} demonstrates that the prevalence of young stellar populations does not indicate ongoing star formation but rather marks a recent cessation of star formation. This finding provides support for evolutionary models where AGN activation occurs after the nuclear starburst phase, positioning AGN as a consequence rather than a catalyst of SF. The discrepancy between studies that find correlations between stellar population age and AGN luminosity versus those that detect no such relationship, particularly with X-ray luminosity measurements, reflects the complexity of isolating cause and effect relationships in these multi-component systems \citep{Burtscher+21}. The fundamental difficulty in resolving this timing debate lies in the challenge of obtaining clean stellar spectra. The necessary technique, stellar population synthesis, requires accurately isolating the host galaxy's stellar continuum from the three competing components present in AGN: the stellar light, the featureless continuum of the accretion disk, and the emission from hot dust. This task is especially challenging in the optical bands due to severe dust reddening and the high luminosity contribution of the AGN accretion disk \citep[]{CidFernandes+04, Riffel+09, Riffel+24}.

To overcome the inherent dust and dilution problems that limit optical studies of AGN, Near-Infrared (NIR $\cong$ $0.8-2.4 \mu$m) spectroscopy provides a superior window. The NIR regime is uniquely sensitive to both the AGN's featureless components and the host galaxy's stellar population features, allowing for their robust disentanglement \citep[]{Riffel+09, Riffel+11c, Riffel+22, Riffel+24, Dahmer-Hahn+18,Dahmer-Hahn+19, Marquez+25}. Crucially, the NIRs' significantly reduced susceptibility to dust attenuation provides a clearer, less biased view of the innermost nuclear stellar populations. This advantage enables reliable stellar population synthesis even for bright Type 1 AGN, where the SMBH completely outshines the stellar light in the optical regime \citep[]{Riffel+09}.

Recent ground-based NIR spectroscopic studies of the LLAMA sample have demonstrated that AGN hosts display higher fractions of intermediate age stellar populations compared to control galaxies, with mass loss rates sufficient to sustain observed AGN luminosities \citep{Riffel+24}. These analyses of larger samples were fundamentally limited by ground-based sensitivity, long-slit, and seeing-limited ($\sim 0.8"$) observations, preventing detailed spatial mapping of stellar population variations and circumnuclear star formation within the inner parsecs.

The power of NIR stellar population analysis has been dramatically enhanced by transformative observations with the Near-Infrared Spectrograph (NIRSpec) on the James Webb Space Telescope (JWST), operating in integral field unit (IFU) spectroscopy mode \citep{Boker+22}. The exceptional combination of outstanding sensitivity (roughly two orders of magnitude better than ground-based adaptive optics-assisted IFUs), high spatial resolution ($\approx 0.1\arcsec$, corresponding to $\approx 45$--$95 \, \mathrm{pc}$ at the distances of our sources), high spectral resolution ($R \cong 2700$),  and the required NIR wavelength coverage ($1.66$--$3.17 \, \mu\mathrm{m}$, our analysis is restricted to the 2.5 $\mu$m spectral region, owing to the current limitations of stellar population synthesis models at longer wavelengths) offered by JWST/NIRSpec provides an unprecedented capability for characterizing the physical conditions within galactic nuclei. These revolutionary observations enable detailed, spatially resolved two-dimensional mapping of the spatial distribution, kinematic structure, and stellar population properties of both gas and stars in the nuclear environment, establishing the NIR as the definitive observational window into AGN host galaxies.

This paper presents a detailed JWST/NIRSpec IFU study of three carefully selected sources: NGC 3884, 3C 293, and CGCG 012-070, from the Blowing Star Formation Away in AGN Hosts \citep[BAH; for details on their selection see][]{Riffel+25,Costa-Souza+24} project. Building upon the optical and ground based NIR stellar population studies \citep[e.g.][]{Riffel+22}, we apply stellar population synthesis using multiple stellar population models: the XSL \citep[X-shooter Spectral Library, ][]{Verro+22}, FSPS \citep[Flexible Stellar Population Synthesis,][]{Conroy+09, Conroy+10}, and M13 \citep[Maraston models,][]{Noel+13} to JWST/NIRSpec data to obtain the first spatially resolved maps of the circumnuclear regions of these AGN.  This paper is organized as follows: Section \ref{sec:data} describes the JWST/NIRSpec observations, target selection, and data reduction procedures. The stellar population synthesis fitting methodology is detailed in Section \ref{sec:fitting}. We present the spatially resolved stellar population analysis and maps in Section \ref{sec:results}. The implications of our findings are discussed in Section \ref{sec:discussion}, followed by the conclusions in Section \ref{sec:conclusions}.

\section{Observations and Data Reduction}\label{sec:data}

\begin{figure}
    \centering
    \includegraphics[width=0.39\textwidth]{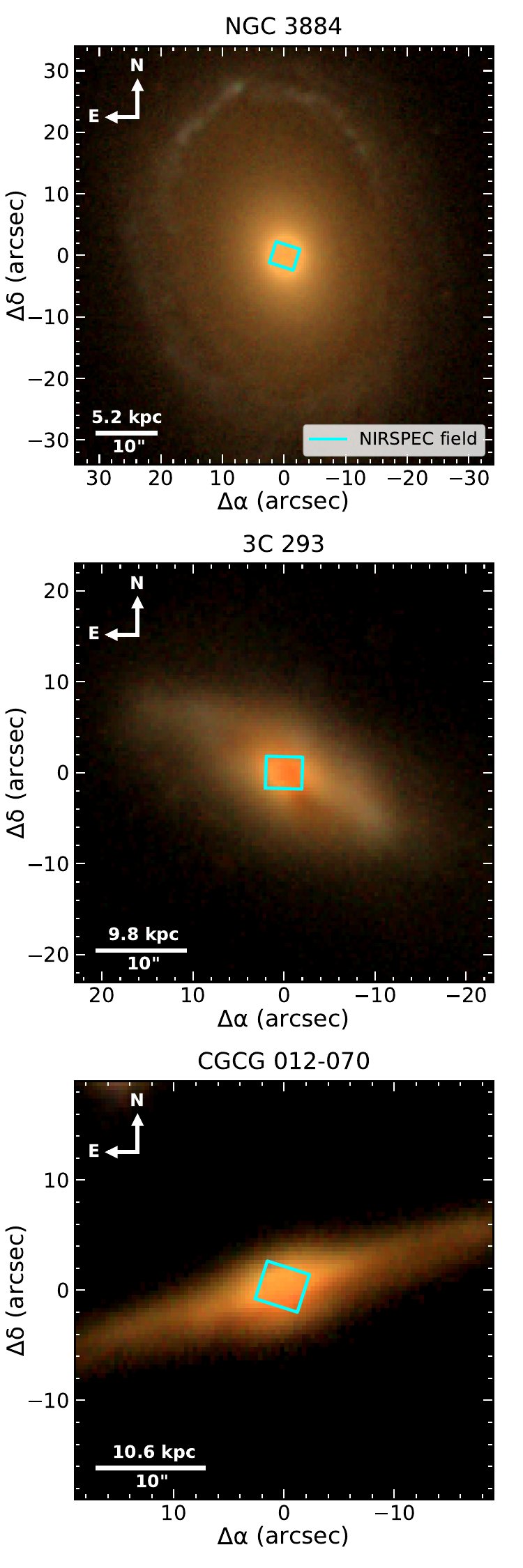}
    \caption{Composite image of the targets in the g (4770\,\AA), r (6231\,\AA), and i (7625\,\AA) bands from the SDSS archive \citep{York+00}. The NIRSpec field of view (FOV) for each galaxy is marked in cyan.}
    \label{fig:sources}
\end{figure}

The galaxies 3C\:293, NGC\:3884 and CGCG~012-070 (Fig.~\ref{fig:sources}) were observed as part of the BAH  \citep[][]{Riffel+25,Costa-Souza+24}  project (PID: 1928, PI: Riffel, R. A.)
using the JWST/NIRSpec instrument in the IFU mode \citep{Boker+22}. These targets selected from \citet{Riffel+20}, corresponding to the sources presenting the most extreme values of the H$_2$~S(3)~9.665\,$\mu$m to PAH~11.3\,$\mu$m ratio, large velocity dispersions in [O\,\textsc{i}]~$\lambda$6300, and elevated [O\,\textsc{i}]~$\lambda$6300/H$\alpha$ line ratios. 
 Such properties indicate that these objects are strong candidates for hosting shocks and molecular gas outflows. Indeed, JWST observations of these sources with MIRI MRS and NIRSpec confirmed the presence of molecular outflows, as well as shock-heated \h2\ emission \citep{Costa-Souza+24,Riffel+25, deMellos+26}.

 The observations were performed with the NIRSpec IFU using the G235H filter in combination with the F170LP grating. We processed the data using the JWST Science Calibration Pipeline \citep{Bushouse+24}, version 1.12.5, and the reference file \texttt{jwst\_1183.pmap}, following the standard pipeline stages. 
We selected a 6-point small-cycling dither pattern and the NRSIRS2RAPID readout mode,   
with integrations comprising 15 groups for NGC3884 and 40 for 3C293 and CGCG012-070.

Additional leakage exposures were obtained and used exclusively for outlier detection in Detector1 stage.  We processed the data using the JWST Science Calibration Pipeline \citep{Bushouse+24}, version 1.12.5, and the reference file \texttt{jwst\_1183.pmap}, following the standard pipeline stages. During the \texttt{Detector1} stage, standard detector corrections and slope fitting were applied. We then used leakage observations to flag hot pixels using a custom method adapted from \citet{Pontoppidan+22}, which identifies high-value pixels along with their neighbors to suppress outliers while preserving real features. The resulting cleaned rate files were passed through the \texttt{Spec2} stage with pixel replacement enabled.

Additionally, we applied NSClean \citep{Rauscher+24} to remove the 1/f noise. In \texttt{Spec3} stage, we tuned the outlier detection parameters (\texttt{kernel="1 11"}, \texttt{threshold=99.5}) to enhance performance, and set the \texttt{output\_type} to "multi" to generate data cubes that combine both detectors, with a spaxel scale width of 0\farcs1. Following this first reduction, we used the generated \texttt{.crf} files in a custom outlier detection script to flag remaining artifacts, updating the data quality masks. The script identifies outliers by recognizing patterns based on intensity contrast with their surroundings. We then repeated the \texttt{Spec3} stage, masking approximately 1–2\% of the science pixels.

Due to spatial undersampling in the NIRSpec IFU, single-spaxel spectra often exhibit wave-like artifacts, commonly referred to as "wiggles" \citep{Perna+23,Law+23}. To mitigate these effects, we adopt a resolution-degradation strategy in which the spectrum of each spaxel is replaced by a continuum-renormalized version derived from the sum of its neighbors' spectra. This method is conceptually similar to applying a Gaussian blur across each wavelength slice, with the renormalization ensuring conservation of the total emitted flux. The procedure effectively reduces or eliminates wiggles, while the flux of the emission lines can vary slightly, typically within the 5–10\% range.

% \input{sections/tabprops}

% FIGURES .... 
% \begin{figure*}
% \begin{minipage}[b]{0.5\linewidth}
% \includegraphics[width=\textwidth]{figures/NGC3081.pdf}
% \end{minipage}\hfill
% \begin{minipage}[b]{0.5\linewidth}
% \includegraphics[width=\textwidth]{figures/NGC2992.pdf}
% \end{minipage}\hfill
% \begin{minipage}[b]{0.5\linewidth}
% \includegraphics[width=\textwidth]{figures/IC4653.pdf}
% \end{minipage}\hfill
% \begin{minipage}[b]{0.5\linewidth}
% \includegraphics[width=\textwidth]{figures/NGC0718.pdf}
% \end{minipage}\hfill
% \caption{Final reduced and redshift-corrected spectra for NGC~3081 (AGN), NGC~2992 (AGN), IC~4653 (CNT) and NGC~0718 (CNT). For each galaxy we show -- the redshift and reddening corrected spectra-- from top to bottom the, $J$, $H$ , and $K$ bands, respectively. The flux is in units of  $\rm 10^{-15}~ erg ~ cm^{-2} ~ s^{-1}$. The shaded grey area represents the uncertainties and the brown area indicates the poor transmission regions between different bands. The remaining spectra are shown in on-line material.}
% \label{spectra}
% \end{figure*}

\section{Stellar Population Fitting}\label{sec:fitting}

\subsection{Fitting code}
Galaxies' spectra are compressed information of various components, including the underlying stellar spectra, gas, and dust emission contributions \citep[][and references therein]{Bica+87,Schmidt+91,Walcher+11,Conroy+13,Gomes+17,Riffel+09,Riffel+11c,Riffel+22}. In the case of active galaxies, additional considerations must be given to components such as the AGN torus and accretion disk \citep{CidFernandes+04, Riffel+09,Burtscher+15,Riffel+22,Riffel+24}. The process of stellar population fitting involves disentangling and determining the percentage contribution of these components to the galaxy spectrum.

Due to the substantial number of parameters involved in characterizing a galaxy's spectrum (e.g., age, metallicity, kinematics, reddening, and AGN components), a plethora of techniques have been continually employed to isolate the distinct spectral constituents \citep[see][for a review]{Walcher+11,Conroy+13}. Consequently, various research groups have developed a multitude of fitting codes, each with its own inherent design priorities \citep[e.g.][]{CidFernandes+05,Ocvirk+06,Koleva+09,Tojeiro+07,Cappellari+17,Sanchez+16,Wilkinson+17,Gomes+17,Johnson+21}. Naturally, inter-code comparisons have been systematically executed, confirming that these codes generally achieve consistent results when applied to the identical data utilizing the same set of models and input parameters \citep[e.g.][]{Koleva+08,Dias+10,Gomes+17,Goddard+17,Ge+18,CidFernandes+18,Woo+24}.

In the present paper, to perform the stellar population fitting, we employ the software \textsc{starlight} \citep{CidFernandes+04,CidFernandes+05,Asari+07,CidFernandes+18}. We opted for this particular choice primarily to ensure consistency with previous studies conducted by our team, enabling easier comparisons of the results. The \textsc{starlight} code fits the complete absorption and continuum features in the observed spectra by combining in different proportions the base-set elements ({\S~\ref{sec:base}). It excludes emission lines and spurious data, employing a blend of computational techniques derived from semi-empirical population synthesis and evolutionary synthesis models \citep{CidFernandes+04,CidFernandes+05}.

In essence, the code fits an observed spectrum, represented as $O_{\lambda}$, using a combination of $N_{\star}$ single stellar populations (SSPs) in varying proportions. The visual extinction ($A_V$) is modeled by \str\ as foreground dust lanes. To fit the JWST data, we use the CCM \citep{Cardelli+89} extinction law. The modeled spectrum, $M_{\lambda}$, is obtained through the following equation:

\begin{equation}
M_{\lambda}=M_{\lambda 0}\sum_{j=1}^{N_{\star}}x_j\,b_{j,\lambda}\,r_{\lambda}  \otimes G(v_{\star},\sigma_{\star})
\label{streq}
\end{equation}
where $x_j$ is the population vector, $b_{j,\lambda}$ is the $j$th base element (see below), $\,r_{\lambda}$ is the reddening factor of the $j$th component normalised at $\lambda_0$, the fitted reddening term is represented by $r_{\lambda}=10^{-0.4(A_{\lambda}-A_{\lambda 0})}$, $M_{\lambda 0}$ is the synthetic flux at the normalisation wavelength (we have used $\rm \lambda_{norm}=21\,940$\AA\ in the rest frame).  The convolution operator is $\otimes$ and $G(v_{\star},\sigma_{\star})$ is the Gaussian distribution used to model the line-of-sight velocity distributions of the stars, which is centered at velocity $v_{\star}$  with dispersion  $\sigma_{\star}$. 
The final fit is carried out through a chi-square minimization, as follows: 

\begin{equation}
\chi^2 = \sum_{\lambda}[(O_{\lambda}-M_{\lambda})w_{\lambda}]^2
\end{equation}
where emission lines and spurious features are excluded from the fit by fixing $w_{\lambda}$=0  at the corresponding wavelengths. 

\subsection{Base of elements}\label{sec:base}

To assess the robustness and internal consistency of our results, we fitted the data using three distinct sets of models. It is important to note that, prior to the fitting procedure, the spectral resolution of either the observational data or the models was adjusted to ensure mutual consistency. Specifically, for the M13 and FPSP fittings, the spectral resolution of the data was degraded to match the lower resolution of the corresponding models, whereas in the XSL case, the models possess a higher intrinsic spectral resolution than the data. All model bases were constructed using different sets of SSPs, each derived from distinct evolutionary population synthesis (EPS) models, as detailed below:

\begin{itemize}

\item[] \textbf{Maraston models (M13):} This base was built following \citet{Riffel+09,Riffel+22} but with the updated \citet{Maraston+05} models as presented in \citet{Noel+13}. The base using these models was built to have 12 ages ($t$=0.01, 0.03, 0.05, 0.1, 0.2, 0.5, 0.7, 1, 2, 5,  9, 13~Gyr)  and 4 metallicities ($Z$ = 0.05, 0.5, 1, 2 $Z_\odot$). They have been selected to have the \citet{Salpeter+55} initial mass function (IMF). These models are based on the fuel consumption theorem and on the use of empirical spectra for C- and O-Rich stars \citep[from][]{Lancon+00}. The hot stars are drawn from the BaSeL theoretical stellar library, in its 2005 version \citep{Lejeune+98}. The M13 models follow a similar prescription to that presented in \citet{Maraston+98,Maraston+05}, however, with a new calibration for the energetics and onset of the TP-AGB phase. For more details, see \citet{Noel+13}.

\item[] \textbf{XSL models (XSL):} One of the bases used was built following \citet{Riffel+24}, using the \citet{Verro+22b} SSPs. These models have been computed with the new X-shooter Spectral Library \citep[XSL,][]{Verro+22} and are fully empirical. The base of elements used here comprises SSPs with 4 metallicities ($Z$ = 0.25, 0.63, 1 and 1.53 $Z_\odot$) and 25 ages ($t$ = 0.050, 0.063, 0.079, 0.1, 0.126, 0.158, 0.200, 0.251, 0.316, 0.398, 0.501, 0.631, 0.794, 1, 1.259, 1.585, 1.995, 2.512, 3.162, 3.981, 5.012, 6.310, 7.943, 10 and, 12.589~Gyr). They have been selected to have \citet{Kroupa+01} IMF and the {\sc parsec/colibri} isochrones \citep{Chen+15,Pastorelli+20}, cover younger ages and are on the models' safe-range \citep[see Fig. 19 of][]{Verro+22b}.

\item[] \textbf{FSPS models (FSPS):} This base was built to be as similar as possible in the age and metallicities parameter space to the M13 models but using the FSPS models \citep{Conroy+09,Conroy+10}.
The base using these models was built to have 12 ages ($t$=0.01, 0.0316, 0.0501, 0.1, 0.2, 0.501, 0.708, 1.0, 2.0, 5.01, 8.91, 12.6~Gyr)  and 5 metallicities ($Z$ = 0.7, 0.85, 1.0, 1.5, 2.0 $Z_\odot$). They have been selected to have the \citet{Salpeter+55} IMF. This model is derived from \citet{Conroy+09, Conroy+10} and uses empirical stellar spectra for the NIR spectral region taken from the NASA InfraRed Telescope Facility (IRTF) stellar library \citep{Cushing+05,Rayner+09} and from its extension \citep[E-IRTF][]{Villaume+17}. For the hot stars, theoretical spectra from the BaSeL3.1 library are used \citep{Lejeune+97,Lejeune+98,Westera+02}. For more details, see \citet{Conroy+09}.

\end{itemize}

Additionally, we followed \citet{Riffel+24} and to account for the accretion disk featureless continuum (FC), we have included in all the different bases three power-laws of the form $F_\lambda \propto \lambda^{-\alpha}$ \citep[e.g.][]{Koski+78,CidFernandes+05,Riffel+09}.The employed values for $\alpha$ are, namely, 0.25, 0.5, and 0.75.  To properly account for the hot dust emission component, eight Planck distributions (black-body, BB), with temperatures ranging from 700 to 1400 K, in steps of 100 K, were included in the fits.  For more details on the definition of these bases, effects of these components in the NIR spectra, as well as possible degeneracies, see \citet{Riffel+09,Riffel+22,Riffel+24}.

Finally, since signal-to-noise effects wash away small differences of individual components, we have binned the population vectors in coarser numbers, but more representative of the fitting \citep[see][for details]{CidFernandes+04,Riffel+09} as follows:
\begin{itemize}
    \item $x{yy} \le 0.05\,\text{Gyr}$;
    \item $0.05\,\text{Gyr} < x{iy} \le 0.20\,\text{Gyr}$;
    \item $0.20\,\text{Gyr} < x{ii} \le 0.70\,\text{Gyr}$;
    \item $0.70\,\text{Gyr} < x{io} \le 2.0\,\text{Gyr}$;
    \item $2.0\,\text{Gyr} < x{o} \le 14.0\,\text{Gyr}$;
    \item FC is the sum over all feature-less components' contributions, and
    \item HD is the sum over all hot dust component contributions.
  
\end{itemize}

For each spaxel, we computed the mean stellar age (more precisely, the logarithm of the age) weighted by stellar light as
\begin{equation}
\langle {\rm log}\, t_{\star} \rangle_{L} = \sum^{N_{\star}}_{j=1} x_j \, {\rm log} t_j,
\end{equation}
and the corresponding mean stellar age weighted by stellar mass as
\begin{equation}
\langle {\rm log}\, t_{\star} \rangle_{M} = \sum^{N_{\star}}_{j=1} \mu_j \, {\rm log} t_j.
\end{equation}
Analogously, the light-weighted mean stellar metallicity is defined as
\begin{equation}
\langle Z_{\star} \rangle_{L} = \sum^{N_{\star}}_{j=1} x_j \, Z_j,
\end{equation}
while the mass-weighted mean stellar metallicity is given by
\begin{equation}
\langle Z_{\star} \rangle_{M} = \sum^{N_{\star}}_{j=1} \mu_j \, Z_j.
\end{equation}

In all cases, these quantities are bounded by the ranges in age and metallicity covered by the adopted base of simple stellar population elements. It should also be noted that the $FC$ and $HD$ components contribute exclusively to the light-weighted properties.

\section{Results}\label{sec:results}
For the Stellar Population fitting of the galaxies, we used three Models: M13, XSL, and FSPS. We focus on results from M13 models for consistency with previous papers by our team \citep[e.g.][]{Riffel+09,Riffel+10,Riffel+11c,Riffel+22} and compare them with those from other models.  

\subsection{NGC~3884}

    NGC~3884 (also catalogued as UGC~06746, CGCG~127-052, CGCG~1143.6+2041, and MCG~+04-28-051) is a spiral galaxy of morphological type SA0/a(r), whose nuclear activity has been classified as Low-Ionization Nuclear Emission-Line Region (LINER)/Seyfert~1 \citep{Veron-Cetty+06}. To the best of our knowledge, we present, for the first time, a stellar population study of this source.
    
    In Figure~\ref{3884M13plot}, we present its spatially resolved stellar population properties derived using the M13 single stellar population (SSP) models. The upper panels display two-dimensional maps of the normalisation flux, the luminosity-weighted and mass-weighted mean stellar age (\maL), the luminosity-weighted mean stellar metallicity (\mzL), and the binned population vectors as defined in \S~\ref{sec:fitting}. The lower panels show spectral synthesis fits for individual spaxels located at representative positions across the galaxy.

    The \maL\ and \mzL\ maps indicate that the galaxy’s emission is dominated by intermediate-age to old stellar populations with elevated metallicities (\mzL\ $\gtrsim 1\,Z_{\odot}$). An analogous behavior is observed in the \maM\ and \mzL\ distributions. In the latter case, however, the maps appear more spatially homogeneous, as expected, given that the mass-to-light ratio of the stellar population components is a non-linear quantity and the AGN emission does not contribute to the stellar mass of the galaxy. This result is further substantiated by the maps of the individual stellar population components. Very young stellar populations ($xyy$ and $xiy$) provide only a minor contribution in the outer regions of the field of view (FoV), with locally enhanced fractions toward the northeastern side at projected galactocentric distances of $\sim$1~kpc from the nucleus. The $xii$ component (0.2~$< t \leq$~0.7~Gyr) is organized in a ring-like configuration (0.4~$\lesssim r \lesssim$~1~kpc), where it contributes approximately 50\% of the observed optical light. This ring surrounds a somewhat older stellar population ($xio$; 0.7~$< t \leq$~2~Gyr), which dominates the emission in the central region and contributes up to $\sim$60\% of the total flux there, whereas the oldest component ($xo$) exhibits a more spatially uniform distribution across the entire FoV.

    The unresolved nuclear continuum emission is dominated by the canonical featureless continuum component ($FC$; $\sim$10\%) and hot dust emission ($HD$; $\sim$30\%). Smaller contributions from these components are detected towards the outer parts of the FoV; these are likely attributable to degeneracies in the spectral fits. For instance, an FC component can be difficult to distinguish from a very young, highly reddened stellar population, and hot dust emission may instead arise from dust distributed in the galactic disk \citep[see][for a detailed discussion]{CidFernandes+95, Riffel+09, Riffel+22, Gaspar+19}.
    
    Comparable results were obtained for NGC~3884 when the XSL models were employed in the data-cube fitting procedure (i.e. a composite stellar population comprising young, intermediate-age, and old components; Fig.~\ref{fig:NGC3884_XSL}). It should be noted, however, that the XSL models do not include stellar populations younger than 50~Myr; consequently, the youngest age bin can only be assigned either to these youngest available templates or to the $FC$. Moreover, the XSL-based fits display small-scale spatial ``structures”, whereas the corresponding maps derived with the M13 models appear smoother.

    In contrast, the fits obtained using the FSPS models do not exhibit a comparable level of agreement. In this case, the stellar light is dominated by the $xyy$ and $xio$ components (each contributing $\sim 50\%$), while the remaining components are predominantly located in the outer regions of the FoV, with $xiy$ being enhanced towards the eastern direction and $xii$ contributing only marginally. The $xo$ component appears as an approximately circular structure at the edges of the FoV. The unresolved nucleus exhibits high fractions of the $FC$ (up to $\sim 30\%$) and $HD$ (up to $\lesssim 20\%$). It is also noteworthy that both the XSL- and FSPS-based solutions present a more ``bumpy” appearance when compared with those obtained using the M13 models, which instead yield smoother solutions consistent with a more continuous SFH. This behaviour is more in line with expectations for spiral galaxies and agrees with the fact that the M13 models are known to perform better for higher-redshift sources when compared with the other two \citep{Lu+24}.

    \figsetstart
\figsetnum{2}
\figsettitle{Spatially resolved stellar population fits for NGC~3884.}
\figsetgrpstart
\figsetgrpnum{2.1}
\figsetgrptitle{NGC~3884 -- M13 models}
\figsetplot{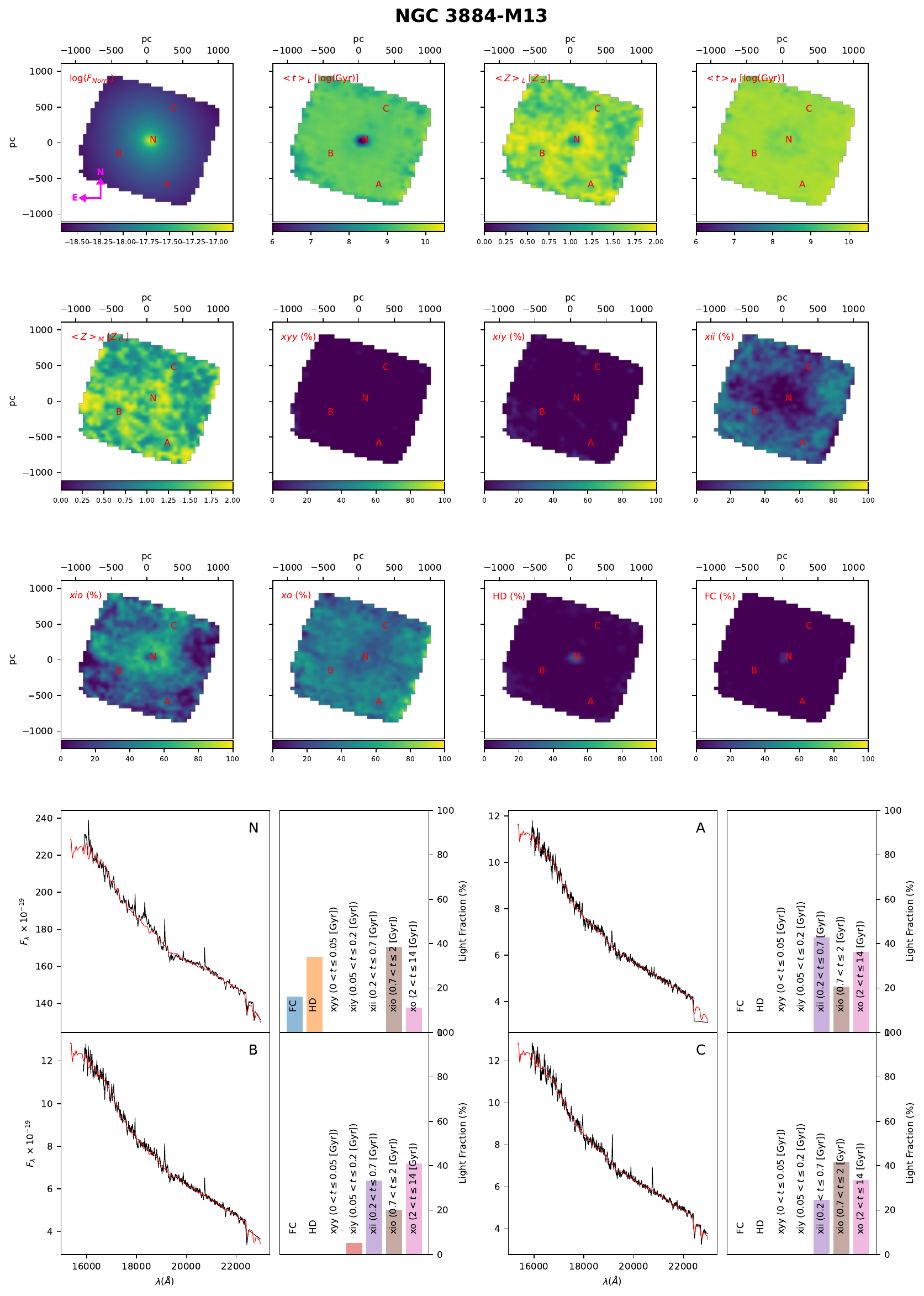}
\figsetgrpnote{Spatially resolved stellar population fits of NGC~3884 using M13 SSP models.}
\figsetgrpend
\figsetgrpstart
\figsetgrpnum{2.2}
\figsetgrptitle{NGC~3884 -- XSL models}
\figsetplot{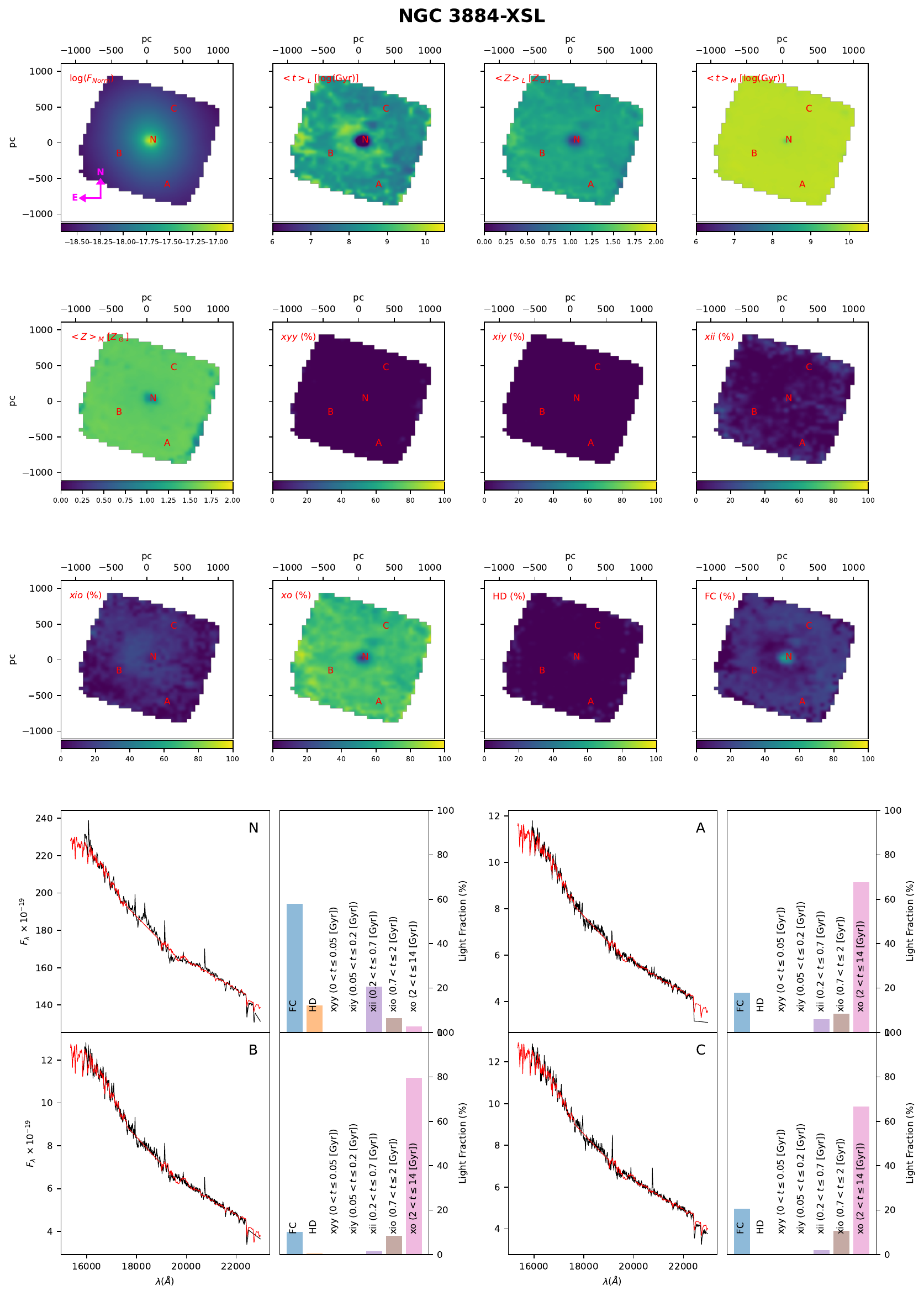}
\figsetgrpnote{Same as Figure~2.1 but for XSL models.}
\figsetgrpend
\figsetgrpstart
\figsetgrpnum{2.3}
\figsetgrptitle{NGC~3884 -- FSPS models}
\figsetplot{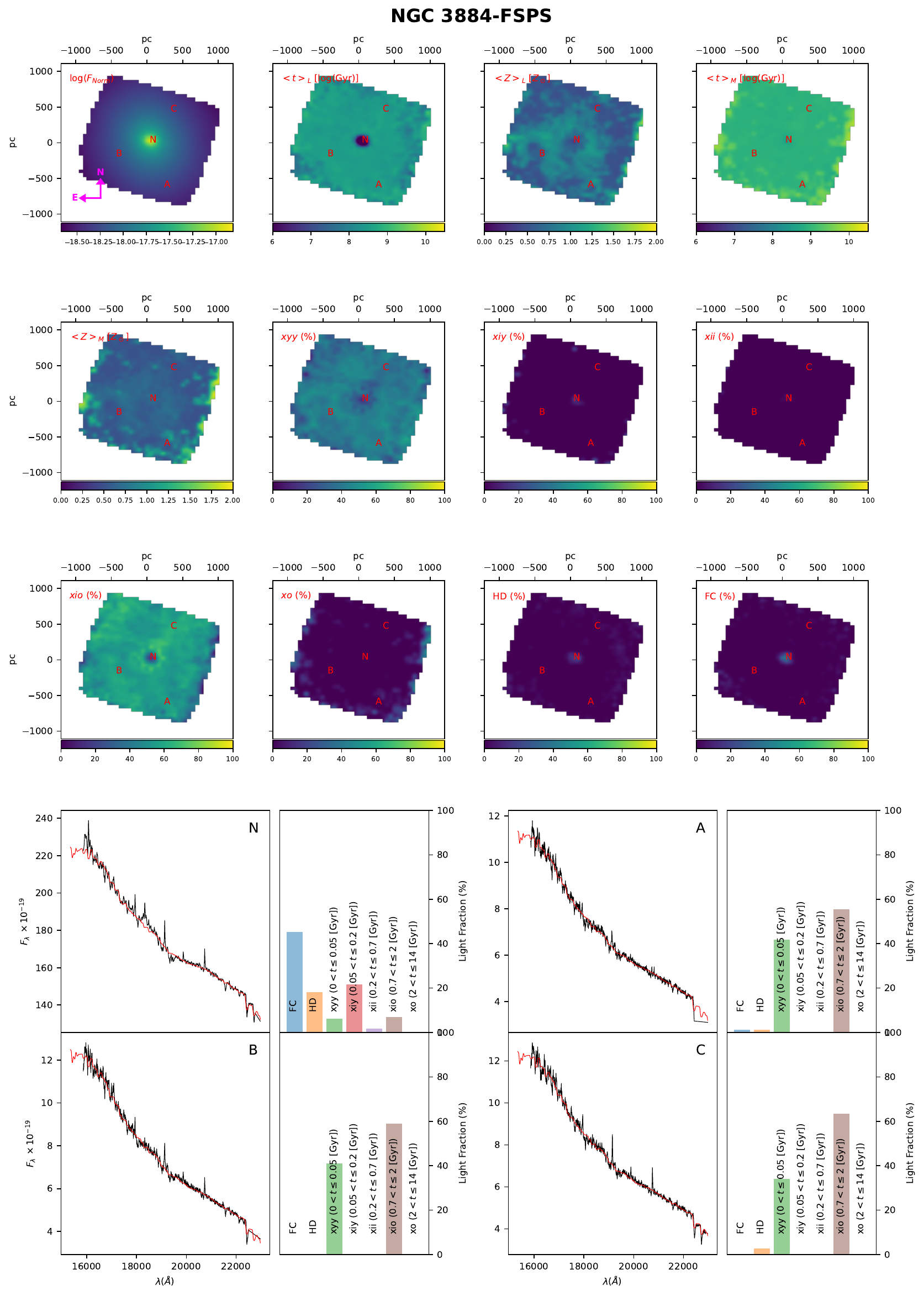}
\figsetgrpnote{Same as Figure~2.1 but for FSPS models.}
\figsetgrpend
%

%%% Static example figure shown in print (NGC 3884 - M13) %%%
    \begin{figure*}
        \centering
        \includegraphics[width=0.85\linewidth]{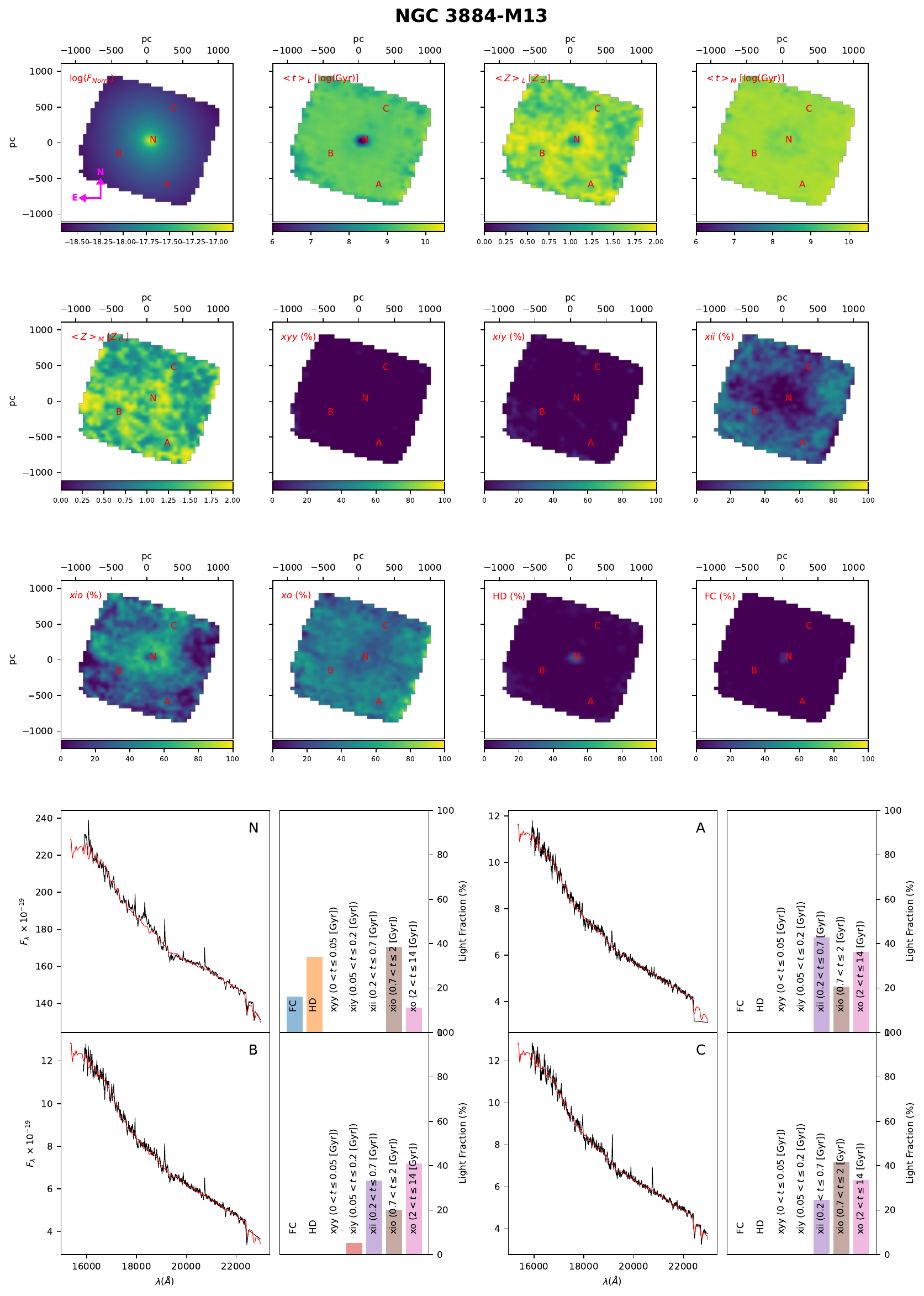}
        \caption{Spatially resolved stellar population fits of NGC~3884 using M13 SSP models. Top panels: Two-dimensional maps of the logarithmic flux (top-left), luminosity-weighted mean age $\langle t \rangle_L$, luminosity-weighted metallicity $\langle Z \rangle_L$, mass-weighted mean age $\langle t \rangle_M$, mass-weighted metallicity $\langle Z \rangle_M$,  and the population vectors as defined in \S~\ref{sec:fitting}
     and labelled in the bottom panels. To enhance the quality of the maps, we applied the Integrated Nested Laplace Approximation (INLA) method \citep{Gonzalez-Gaitan+19}. The labels N, A, B, and C denote individual spaxels used to illustrate detailed fits in the bottom panels. In each panel, the observed spectrum is plotted in black, with the best-fitting model overlaid in red. The histograms show the star-formation history, with the contributions of the spectral components to the total light at the normalization wavelength. The complete figure set (3 images) is available in the online journal.}
        \label{3884M13plot}
        \label{fig:NGC3884_XSL}
        \label{fig:NGC3884_FSPS}
    \end{figure*}

\subsection{3C~293}

The source 3C~293 (also catalogued as UGC 08782, VV 369, CGCG 162-021, CGCG 1350.0+3142, and MCG +05-33-012) is classified as a spiral galaxy \citep{deVaucouleurs+91} hosting a LINER nucleus \citep{Veron-Cetty+06}. To our knowledge, no detailed characterization of the stellar content of this radio galaxy has been reported in the literature; we therefore present below the results of our analysis.

The maps of 3C~293 fitted with the M13 models are displayed in Fig.~\ref{fig:3C_293_M13}. Analogously to the case of NGC~3884, we provide representative examples of spectral fits at distinct locations across the galaxy. Examination of the \maL\ (\maM) and \mzL\ (\mzM) maps indicates that the stellar content is dominated by intermediate-age to old populations with high metallicity (\mzL $ \gtrsim 1\,Z_{\odot}$). The very young component ($xyy$) contributes only marginally, primarily in the outer regions of the field of view (FoV), with modest enhancements toward the northeast at a projected distance of approximately 1~kpc from the nucleus. The $xiy$ component is not detected in the nuclear region but appears in a ring-like configuration (0.5 $\lesssim r \lesssim$ 1.2\,kpc), where it exhibits localized enhancements in the southeast direction, reaching contributions of up to $\sim 30\%$.

The $xii$ component is distributed over the entire FoV, contributing up to $\sim$50\% of the flux in these regions; however, there is a ``hole" located at the nucleus where this component drops to zero. The $xio$ component is also observed over the entire FoV, showing enhancements in locations close to the N and C spaxels, where it reaches up to 60\% of the flux. The underlying old component ($xo$) is distributed across the entire FoV, with contributions between $\sim$30\% and $\sim$ 60\%. Interestingly, the $HD$ contribution is centred on the nucleus (at the peak of the continuum $F_{norm}$) and is elongated along the N-S direction. Finally, the $FC$ component appears unresolved in two locations (N and B), suggesting that this galaxy could host a double AGN.

The fits using XSL models (Fig.~\ref{fig:3C_293_XSL}) are in agreement with those of M13; however, the contribution of the two younger components is negligible, while the $xii$ contribution is small ($\sim$10\%), being enhanced over the SE direction. The dominant stellar population across the entire FoV is $xo$, reaching up to 75\% in some locations. Hot dust emission has a very similar morphology to that seen when fitting M13 models; $FC$ is also enhanced in the N and B locations; however, smaller fractions $\lesssim$20\% are required in the other FoV locations. Since the young populations and the $FC$ are degenerated \citep[e.g.][]{Riffel+22,Riffel+23,Riffel+24}, this may indicate that, according to these fits, a younger population than the model's limit ($t<$50\,Myr) would be required to properly fit the underlying continuum of this source and, in that case, it is represented by the $FC$.

Conversely, when employing the FSPS models (Fig.~\ref{fig:3C_293_FSPS}), the very young component $xyy$ is detected throughout the entire FoV, reaching fractional contributions of up to 50\% in certain regions (e.g., the C spaxel). The contributions from $xiy$ and $xii$ are negligible, whereas the presence of $xio$ is required across the whole FoV, reaching up to 50\% in some locations (e.g., the A spaxel).

    \figsetstart
\figsetnum{3}
\figsettitle{Spatially resolved stellar population fits for 3C~293.}

\figsetgrpstart
\figsetgrpnum{3.1}
\figsetgrptitle{3C~293 -- M13 models}
\figsetplot{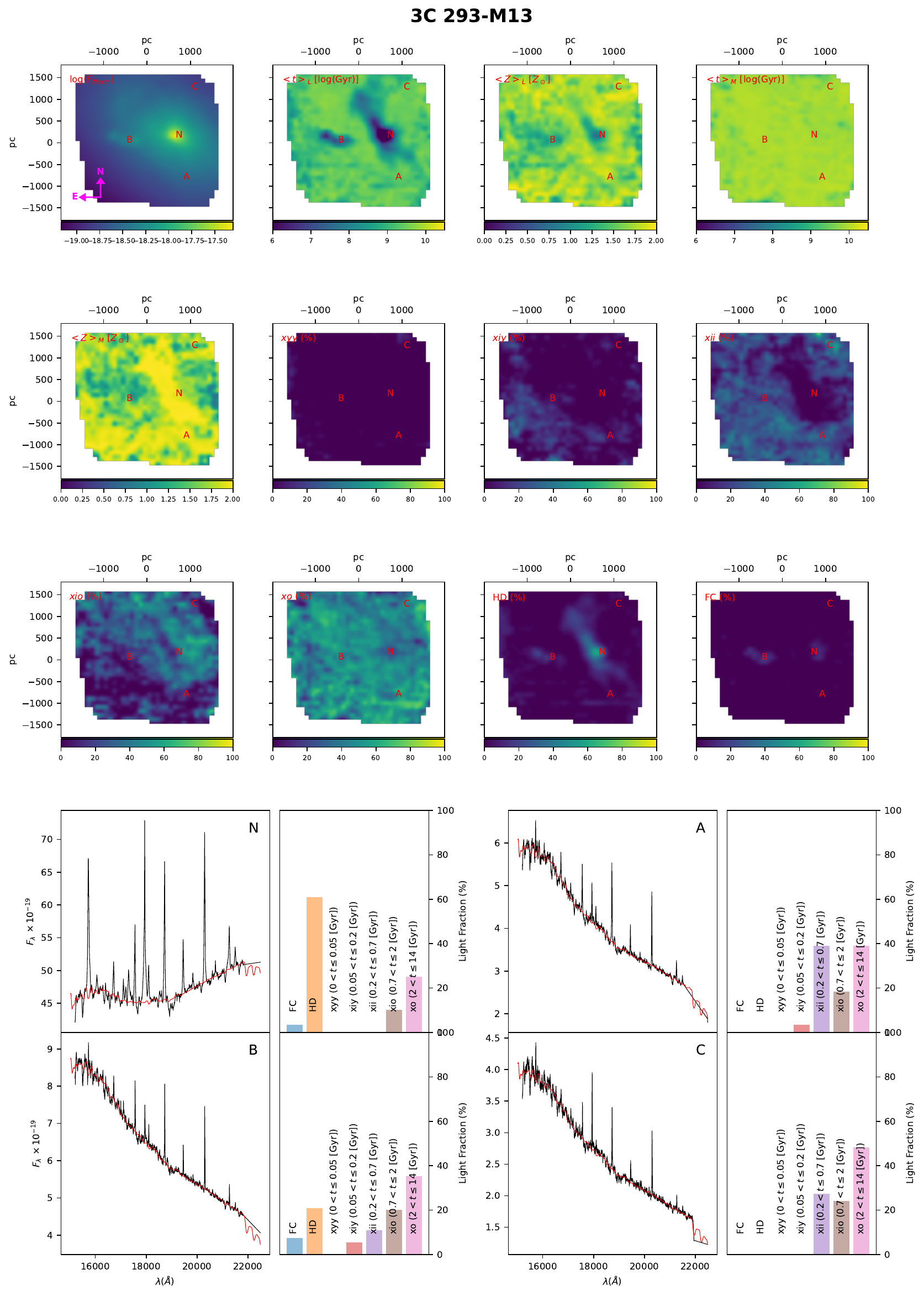}
\figsetgrpnote{Same as Figure~2.1 but for 3C~293.}
\figsetgrpend
\figsetgrpstart
\figsetgrpnum{3.2}
\figsetgrptitle{3C~293 -- XSL models}
\figsetplot{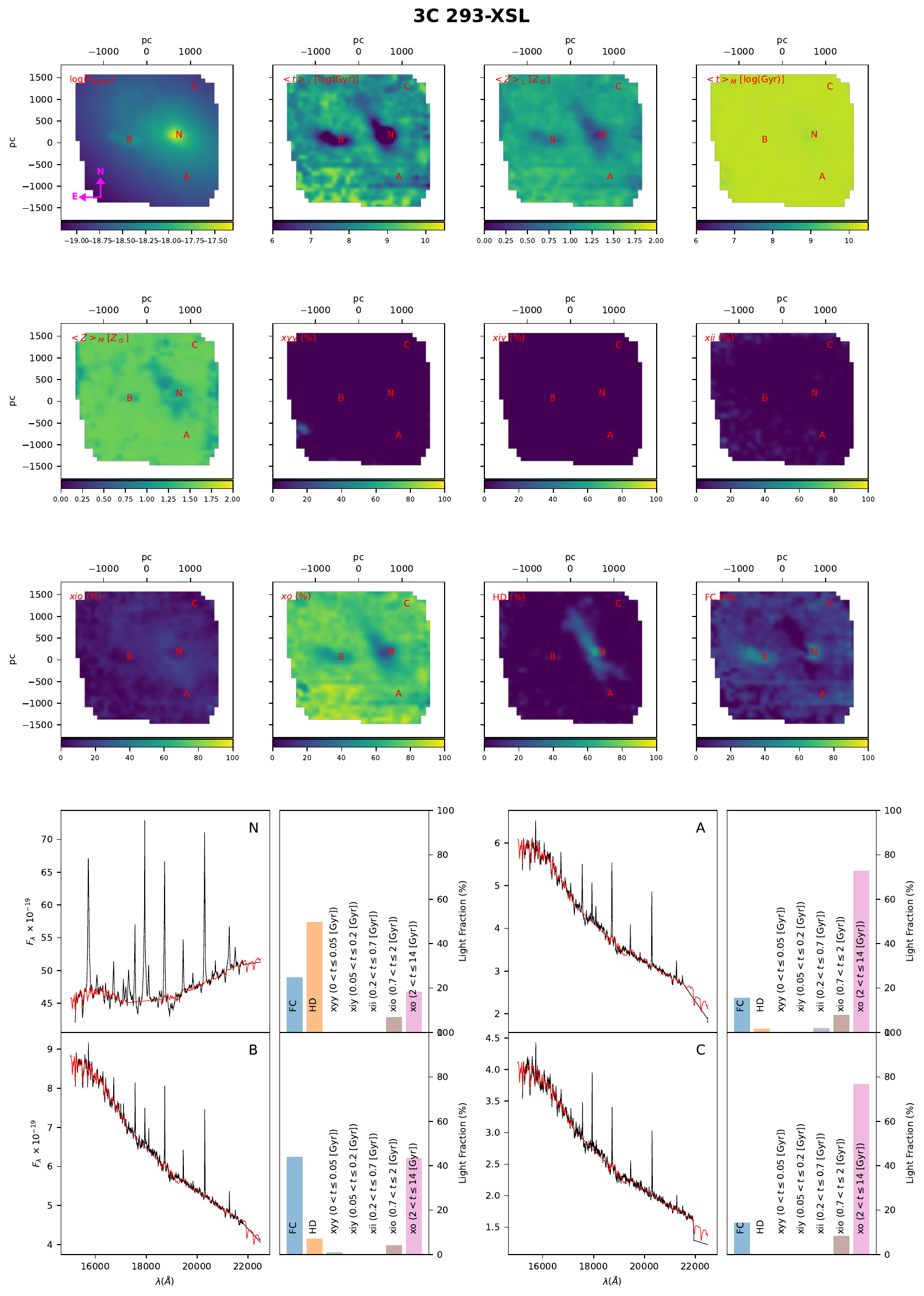}
\figsetgrpnote{Same as Figure~2.1 but for XSL models and 3C~293.}
\figsetgrpend
\figsetgrpstart
\figsetgrpnum{3.3}
\figsetgrptitle{3C~293 -- FSPS models}
\figsetplot{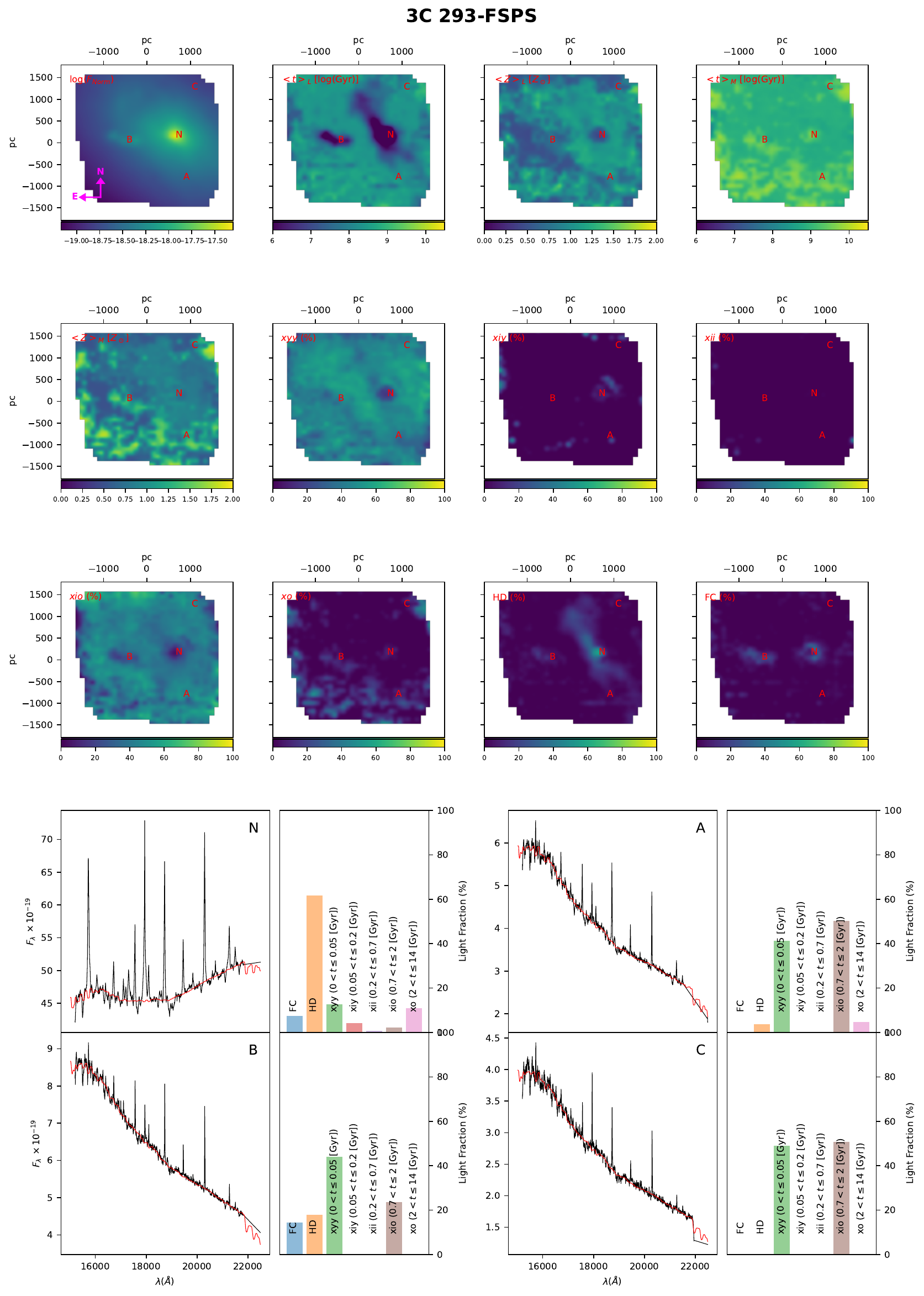}
\figsetgrpnote{Same as Figure~2.1 but for FSPS models and 3C~293.}
\figsetgrpend
\figsetend

%%% Static example figure shown in print (3C - M13) %%%

\begin{figure*}
   \centering
   \includegraphics[width=0.85\linewidth]{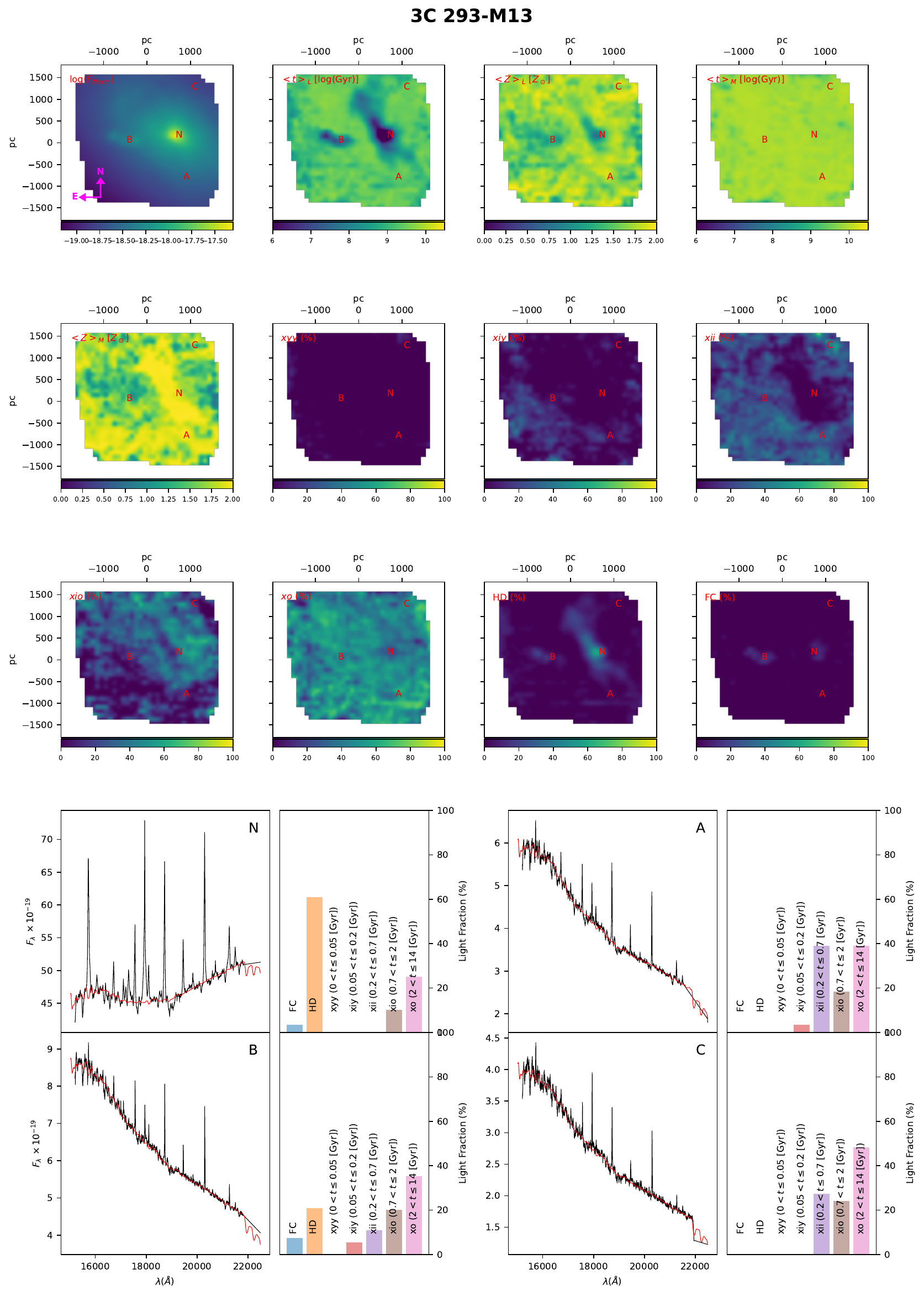} 
   \caption{Same as Figure \ref{3884M13plot} but for 3C 293.
   The complete figure set (3 images) is available in the online journal.}
        \label{fig:3C_293_M13}
        \label{fig:3C_293_XSL}
        \label{fig:3C_293_FSPS}

\end{figure*}

\subsection{CGCG~012-070}

This galaxy (with cross-identifications CGCG 1142.0-0317, 2MFGC 09214, WISEA J114428.78-033415.3, and 2MASX J11442881-0334156) hosts a Seyfert nucleus \citep{Veron-Cetty+06} and is morphologically classified as an Sa system \citep{Nair+10}. As for the other objects in our sample, despite its presence in multiple catalogs, there is little to no information available in the literature regarding its stellar content. 

Figure \ref{fig:CGCG012_M13} displays the spatially resolved parameter maps and representative spectral fits for CGCG~012-070 obtained using the M13 models, and reveals a system that differs from the previously analyzed galaxies. The luminosity-weighted mean stellar age and metallicity maps (see also the mass-weighed ones) indicate an old and chemically evolved stellar population ($t \gtrsim 5$ Gyr; $Z \gtrsim 1.5 Z_{\odot}$), which is remarkably homogeneous across the entire FoV. This is consistent with the fractional light contributions of the individual stellar population components: the $xyy$ component is negligible, while $xiy$ contributes only modestly (typically $\lesssim 10\%$). The remaining three components are present throughout the FoV, with $xii$ reaching $\sim 20\%$ in a few localized regions and $xio$ contributing up to $\sim 40\%$ (e.g., in spaxel B). The $xo$ component is dominant at all positions, with a fractional contribution $\gtrsim 20\%$. Finally, no contribution from the $FC$ component is detected anywhere within the FoV, whereas an $HD$ continuum is required to reproduce the spectra of a small number of spaxels forming a “stream”-like structure toward the south. Even in these regions, however, the $HD$ component accounts for only a minor fraction of the spectral energy distribution ($\lesssim 5\%$).

Comparable results are obtained when fitting the data with the XSL models (Fig.~\ref{fig:CGCG012_XSL}). In this case, however, the old stellar population component $xo$ provides the dominant contribution, whereas the intermediate-age components $xii$ and $xio$ account for smaller fractions of the total light, and the two youngest components ($xyy$ and $xiy$) are not required by the fit. The same stream-like structure toward the south is recovered in the $HD$ component. Nonetheless, similarly to what is observed for 3C~293, a small but non-negligible ($\sim 10\%$) contribution from the $FC$ component is detected across the entire FoV.

In contrast, when the FSPS models are employed (Fig.~\ref{fig:CGCG012_FSPS}), the very young component $xyy$ is detected throughout the full FoV, reaching fractional contributions of up to 60\% in specific regions (e.g., the C spaxel). The contributions from $xiy$ and $xii$ are negligible, while the presence of $xio$ is required across the whole FoV, attaining values of up to 50\% in some locations (e.g., the B spaxel). Only a very small fraction of $xo$ is needed to reproduce the spectra of spaxels in the E direction. No $FC$ contribution is required in these fits, and the stream-like structure remains evident in the $HD$ emission.

    \figsetstart
\figsetnum{4}
\figsettitle{Spatially resolved stellar population fits for CGCG~012-070.}
\figsetgrpstart
\figsetgrpnum{4.1}
\figsetgrptitle{CGCG~012-070 -- M13 models}
\figsetplot{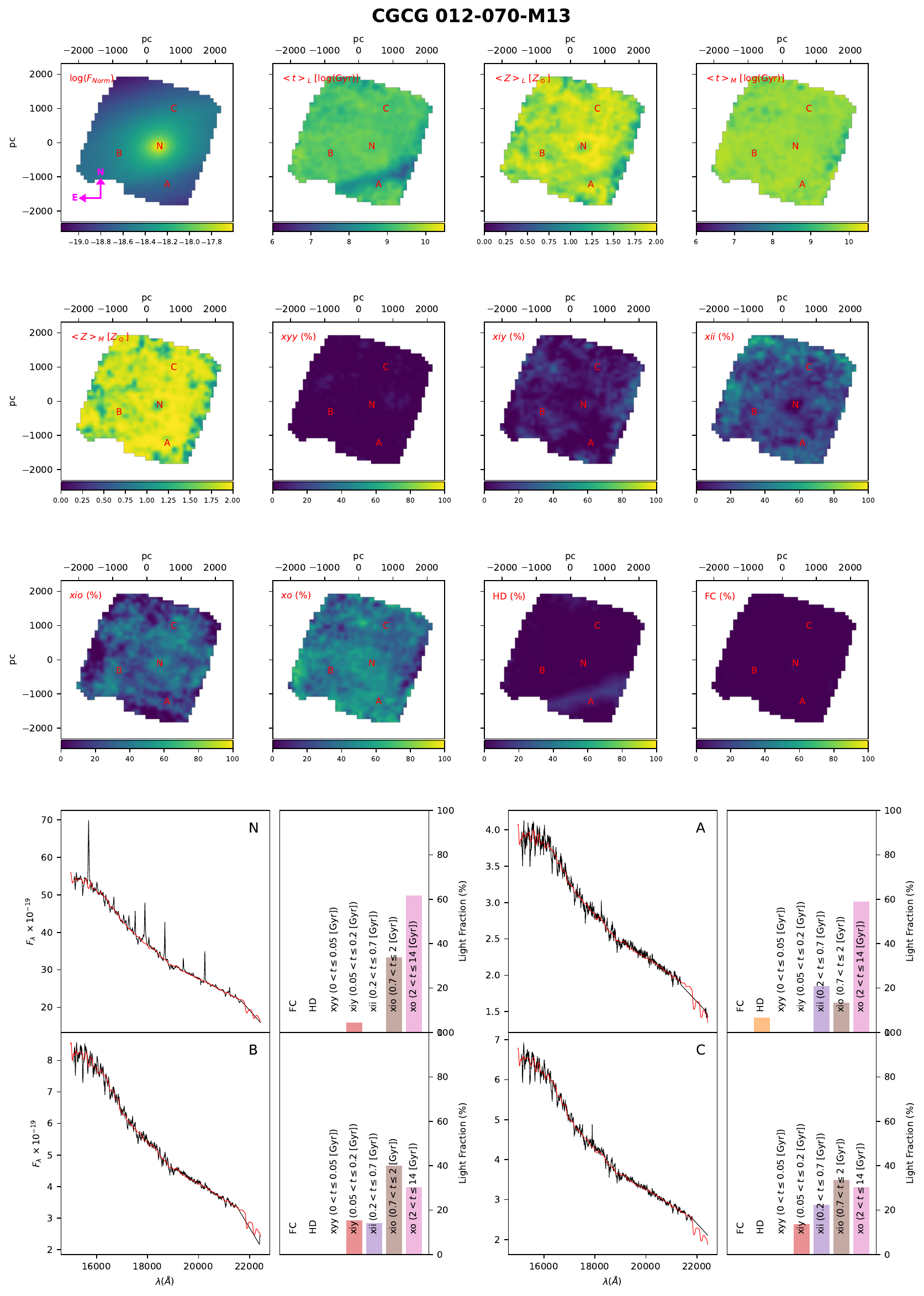}
\figsetgrpnote{Same as Figure~2.1 but for CGCG~012-070.}
\figsetgrpend
\figsetgrpstart
\figsetgrpnum{4.2}
\figsetgrptitle{CGCG~012-070 -- XSL models}
\figsetplot{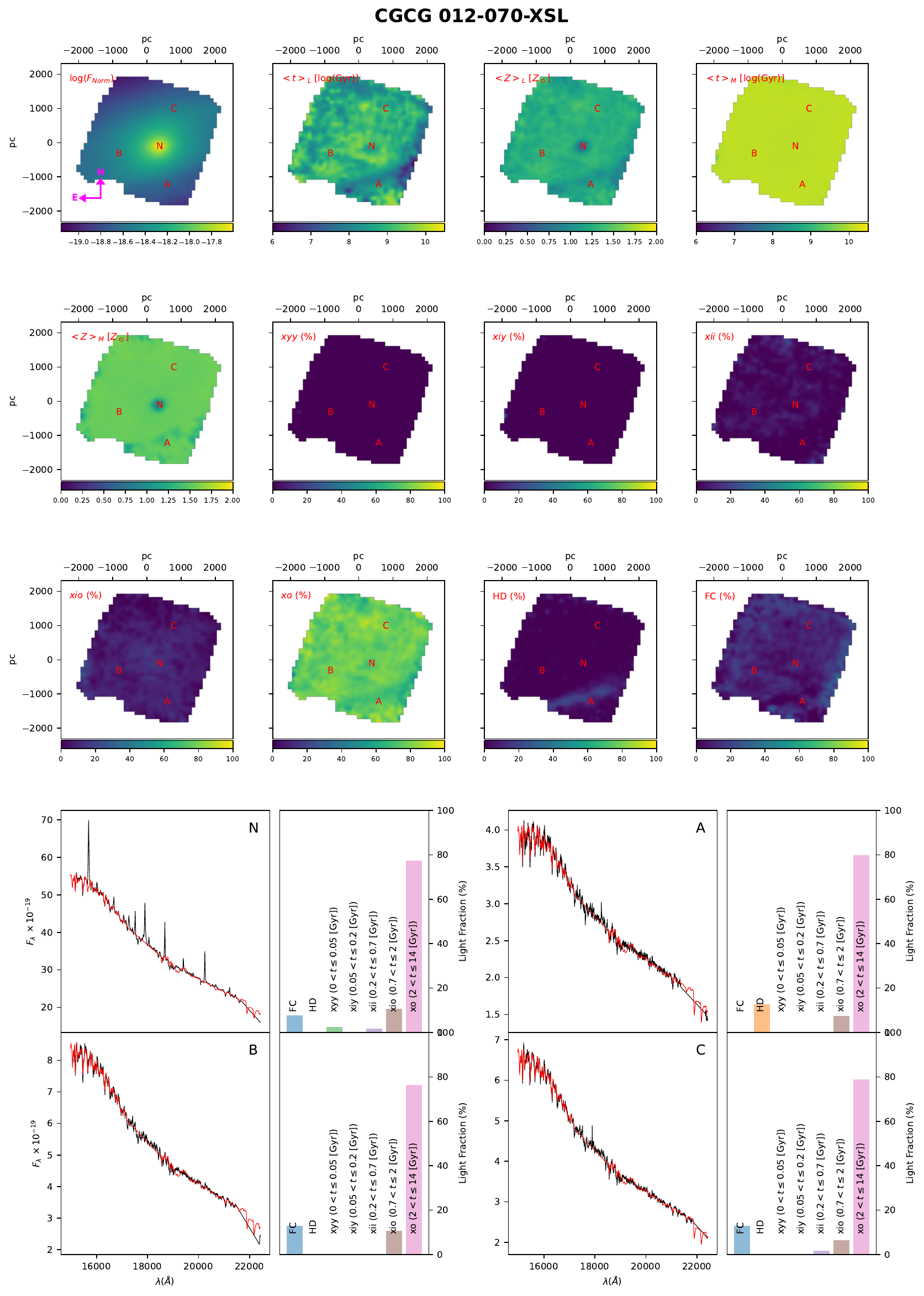}
\figsetgrpnote{Same as Figure~2.1 but for XSL models and CGCG~012-070.}
\figsetgrpend
\figsetgrpstart
\figsetgrpnum{4.3}
\figsetgrptitle{CGCG~012-070 -- FSPS models}
\figsetplot{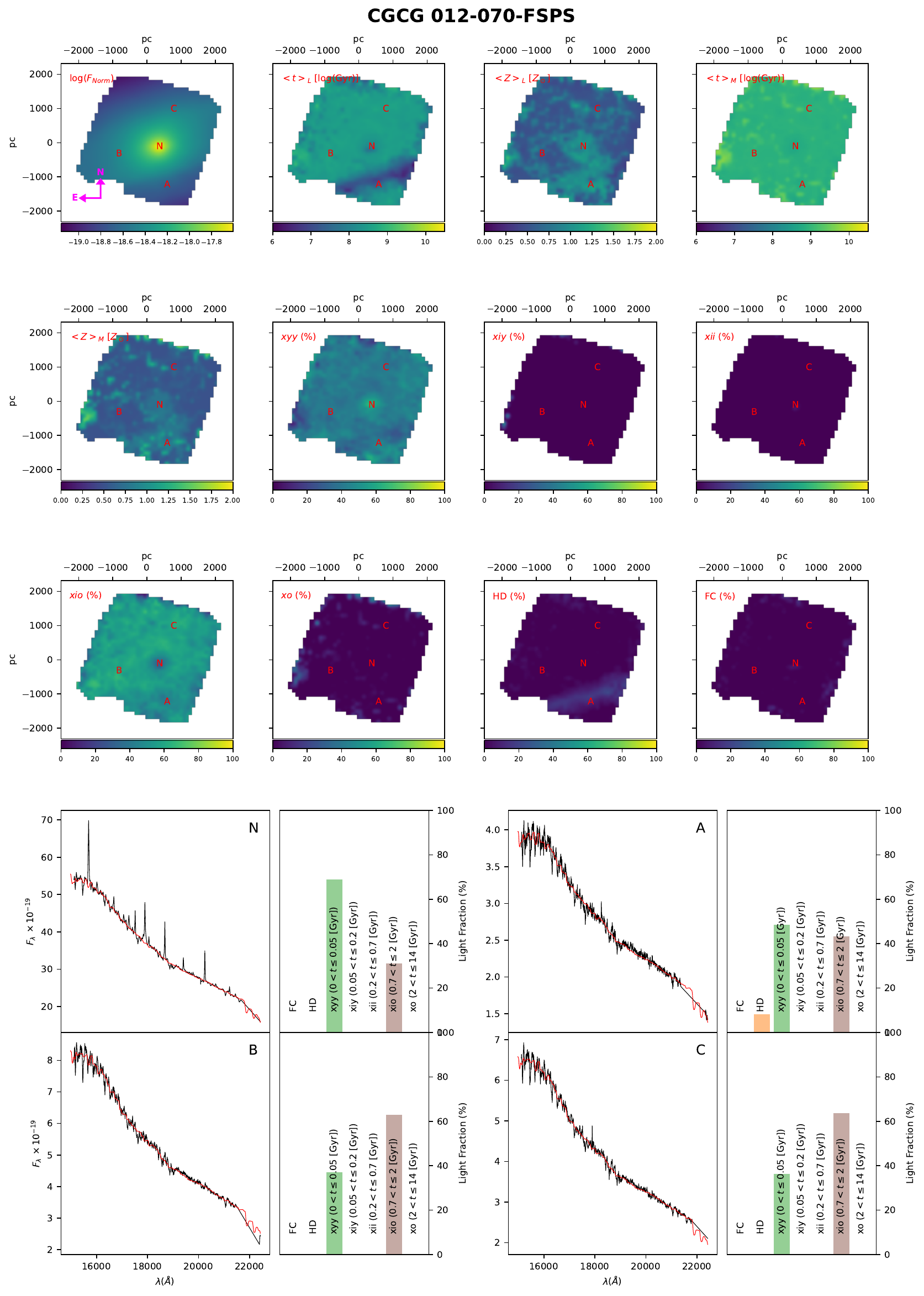}
\figsetgrpnote{Same as Figure~2.1 but for FSPS models and CGCG~012-070.}
\figsetgrpend
\figsetend

%%% Static example figure shown in print (3C - M13) %%%

\begin{figure*}
   \centering
   \includegraphics[width=0.85\linewidth]{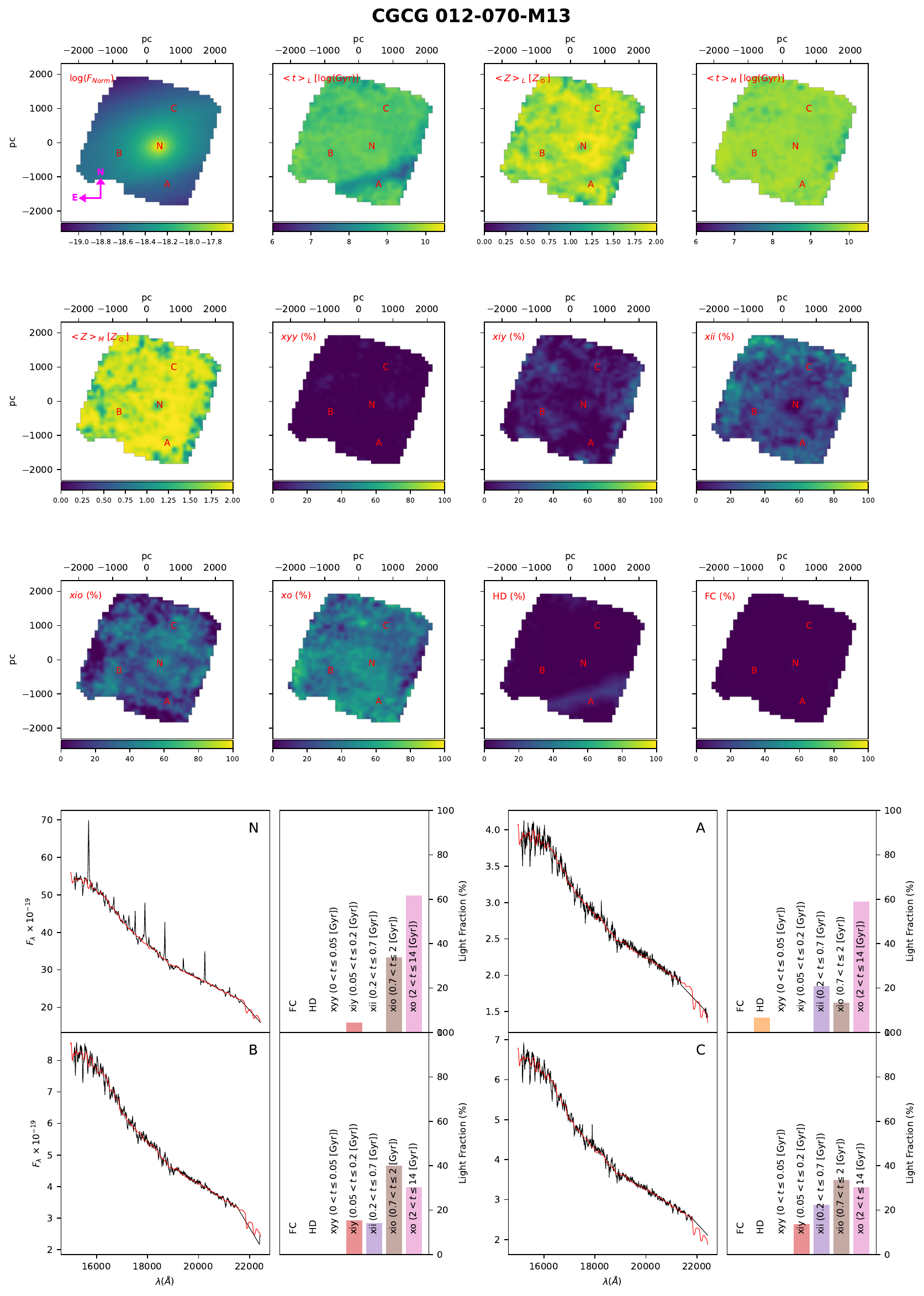} 
   \caption{Same as Figure \ref{3884M13plot} but for CGCG~012-070.
   The complete figure set (3 images) is available in the online journal.}
        \label{fig:CGCG012_M13}
        \label{fig:CGCG012_XSL}
        \label{fig:CGCG012_FSPS}

\end{figure*}

\section{Discussion}\label{sec:discussion}

\subsection{Individual sources}

\textbf{NGC 3884:} The nuclear activity of this galaxy is classified as a LINER/Sy~1, characterized by weak H$\alpha$ emission originating in the broad-line region \citep{Veron-Cetty+06,Riffel+24a}. The ionized and neutral gas kinematics within the inner $\sim 1.4\,\mathrm{kpc}$ of NGC 3884 reveal three distinct components: (i) a disk component whose kinematics are consistent with those of the stellar component; (ii) an inflowing gas component, or gas potentially acquired externally through past merger events, inferred from a twist in the kinematic position angle of the gaseous velocity field in the inner $\sim 340\,\mathrm{pc}$; and (iii) an outflow component, identified by broad emission-line wings ($\sigma \sim 250$–$400\,\mathrm{km\,s^{-1}}$), with a maximum mass outflow rate of $0.25 \pm 0.15\,M_{\odot}\,\mathrm{yr^{-1}}$ \citep{Cazzoli+18,HermosaMunoz+22,Riffel+24a}. In addition, this galaxy hosts a weak radio source with a Faint Images of the Radio Sky at Twenty Centimetres (FIRST) 1.4 GHz luminosity of $L_{\mathrm{FIRST}} \approx 1.01 \times 10^{22}\,\mathrm{W\,Hz^{-1}}$ \citep{Lofthouse+18}. 

Using GMOS data, \citet{Riffel+24a} studied the kinematics of this source and reported an ionized-gas outflow, as well as a misalignment between the gaseous and stellar disks, which may be indicative of a recent interaction and the trigger of the AGN \citep{Raimundo+23,Raimundo+25}. In \citet{Riffel+25} we identified a nuclear ring detected in PAH emission. This ring was also detected in $\rm H_2$ emission (Costa-Souza, in preparation). The above results would typically suggest the presence of young stars (which, in that scenario, would provide the ionizing photons responsible for the PAH and H$_2$ emission). However, as shown in Fig.~\ref{3884M13plot}, we do not detect any significant contribution from very young stellar populations. Instead, a ring co-spatial with the molecular emission is observed in the intermediate-age component $x_{io}$ ($0.70\,\text{Gyr} < x_{io} \le 2.0\,\text{Gyr}$) and on the \maM\ map. A plausible interpretation is that the molecular gas in the nuclear region is currently being dissociated and excited by the AGN radiation \citep[e.g.][]{Sales+10,Riffel+13}, which is suppressing ongoing star formation in this source.

{\bf 3C~293} is a radio-loud LINER \citep{Veron-Cetty+06,Riffel+23b}. Optical imaging reveals a complex morphology characterized by filamentary dust lanes extending from several hundred parsecs up to kiloparsec scales, most likely indicative of a past merger event \citep{Martel+99}. The host galaxy contains a double-double radio source, with outer lobes extending to $\sim$200 kpc and oriented approximately NW–SE, and inner lobes of $\sim$4 kpc in extent aligned roughly E–W \citep{Machalski+16}. 

Nuclear outflows have been detected in neutral hydrogen, reaching velocities of $\sim$1400 km s$^{-1}$ from the western radio hotspot \citep{Morganti+03,Mahony+13}, and in ionized gas, with velocities up to $\sim 1000\,\mathrm{km\ s^{-1}}$ co-spatial with the eastern radio knot, located opposite to the neutral outflow region \citep{Mahony+13}. These outflows are generally interpreted as signatures of strong jet–interstellar-medium (ISM) interaction. Using multifrequency radio observations from LOFAR, MERLIN, and the VLA, \citet{Kukreti+22} mapped the spectral index distribution and demonstrated that the inner lobes represent a young, jet-dominated source undergoing strong interaction with the ISM, likely responsible for driving the ionized outflows. A comparison of the spectral indices of the inner and outer structures indicates that 3C~293 has experienced at least two distinct episodes of nuclear activity. 

\citet{Riffel+23b} showed that the kinetic coupling of the outflows is sufficient to significantly impact the star formation efficiency in this system. \citet[][see also \citealt{Costa-Souza+24}]{deMellos+26} reported that the kinematic center is offset from both the optical and infrared nuclei, which they interpreted as further evidence of a past merger event in this galaxy \citep{Floyd+06}. They also found that 3C~293 exhibits exceptionally high gas extinction and identified a secondary Pa$\alpha$ emission blob east of the nucleus, for which they could not provide a unique explanation, suggesting it could be associated with either star formation or a secondary AGN. 

This secondary Pa$\alpha$ feature coincides with our B position in Fig.~\ref{fig:3C_293_M13}. At this location, we also identify a second $FC$ peak, which lends support to the hypothesis that 3C~293 hosts a secondary AGN, separated from the primary by only $\sim$1 kpc. Such a small projected separation would place this system at the extreme, and therefore very rare, end of the known dual-AGN population \citep[e.g. dual AGN constitute $\sim$4 per cent of the overall AGN population; for projected separations between 4 and 30 kpc, those dual systems residing within a single galaxy represent only $\sim$15 per cent of all duals and exhibit a mean separation of $\sim$10 kpc]{Volonteri+22}. Nonetheless, because the contributions from the $FC$ and ionising stellar populations are degenerate \citep[e.g.][]{CidFernandes+04,Riffel+09}, a more detailed analysis is required to place robust constraints on the presence and properties of the putative secondary AGN in this source.

{\bf CGCG~012-070:}  The nuclear activity of this galaxy is classified as Seyfert~2 \citep{Veron-Cetty+06}. It exhibits weak radio emission, with a 1.4~GHz monochromatic radio power of \(P_{1.4} \approx 2.4 \times 10^{22}\ \mathrm{W\,Hz^{-1}}\), as measured by the FIRST survey \citep{Lofthouse+18}. \citet{RamosVieira+25} investigated the ionized gas kinematics, traced by strong emission lines, and identified two distinct components: a narrow component ($\sigma \lesssim 200\,\mathrm{km\,s^{-1}}$) confined to the galactic disc and co-rotating with the stellar component, and a broad component ($\sigma \gtrsim 300\,\mathrm{km\,s^{-1}}$) associated with an outflow within the inner $\sim 1\,\mathrm{kpc}$. The emission from the disc component is predominantly powered by photoionization from the active galactic nucleus (AGN), whereas the outflowing component additionally includes shock-excited gas, as inferred from emission-line flux ratio diagnostics. The outflows are consistent with being radiatively driven, with a mass outflow rate of $(0.067 \pm 0.026)\,M_\odot\,\mathrm{yr^{-1}}$ and a kinetic coupling efficiency of $0.07$ per cent, and may contribute to the redistribution of gas and the operation of maintenance-mode AGN feedback in CGCG~012-070. 
\citet{Riffel+25a} report the detection of a galaxy gravitationally lensed by CGCG~012-070. The background source is identified through the spatial distribution of rest-frame optical emission-line fluxes, consistent with a redshift of $z \sim 2.89$. The system is detected in [O~III]~$\lambda\lambda 4959, 5007$, H$\beta$, and H$\alpha$ emission lines, and exhibits line ratios characteristic of a star-forming galaxy. The emission-line flux distributions reveal three distinct components, which are modeled using an elliptical power-law mass profile for the lens galaxy. However, as can be seen in Fig.~\ref{fig:CGCG012_M13} these kinematical and lensed galaxy have not left clear fingerprints on the stellar population maps of this source.

\subsection{General results}
Following \citet{Riffel+22}, to investigate the radial behavior of the star formation history (SFH) along the galaxies, we computed, for each object, the mean fraction contribution of the stellar population components within four radial bins ($0 < r \leq 500$ pc [red], $500 < r \leq 1000$ pc [blue], $1000 < r \leq 1500$ pc [green], and $1500 < r \leq 2000$ pc [magenta], centered at the peak of the continuum). We then averaged these quantities over all galaxies in the sample to obtain a representative radial SFH profile for the BAH sample. The resulting distributions are presented in Fig.~\ref{fig:combinedM13} and display the same general behaviors discussed in \S\ref{sec:results}. The analysis based on the XSL models (Fig.~\ref{fig:combinedXSL}) yields qualitatively similar results to those obtained with the M13 models, namely a complex SFH with significant contributions from all stellar population components across the entire FoV. The SFH recovered using the FSPS models (Fig.~\ref{fig:combinedFSPS}) is, on the other hand, strongly dominated by the $xyy$ and $xio$ components, with negligible or very low contributions from the remaining components, thus highlighting even more the fact that the central regions of these galaxies have a sizable amount of young to intermediate age stars. Finally, while the XSL and FSPS frameworks tend to yield more irregular, “spiky” SFH solutions, the M13-based analysis produces smoother and more continuous SFHs, which is more expected in galaxies and reflects results found for higher redshift sources \citep{Lu+24}.

Despite the limited size of the sample, several noteworthy trends emerge. The average radial behavior is broadly consistent with that derived from the Gemini Near-Infrared Integral Field Spectrograph (NIFS) adaptive-optics-assisted data cubes analyzed in \citet{Riffel+22}, revealing two distinct spatial patterns: the $HD$ and $FC$ components decrease with increasing radius, whereas the younger stellar components (up to $xii$) show an outward increase. Conversely, the $xio$ component declines with radius, while the $xo$ component remains approximately constant throughout the FoV. These results are in agreement with the findings of \citet{Rowlands+18} for inactive galaxies of comparable stellar mass. 

Our results can be directly compared with those reported by Marinho et al. (submitted), who investigated a sample of 26 BASS Seyfert~2 galaxies based on optical MUSE observations. Their analysis revealed a significant contribution from young stellar populations in the nuclear regions, with an average light fraction of approximately $20\%$, accompanied by an intermediate-age component contributing around $10\%$. Both the young stellar component and the featureless continuum ($FC$), which attains average central values of roughly $10\%$, exhibit a marked radial decline, whereas the intermediate-age component displays the opposite trend, increasing toward larger galactocentric distances. Furthermore, \citet{Armah+23} reported a decrease in the gas-phase metallicity in the central regions of AGN hosts due to the inflow of metal-poor gas \citep[see also][]{doNascimento+22}. Similarly, Marinho et al. (submitted) identified a decline in stellar metallicity toward the center, suggesting that this pristine gas fuels recent star formation and imprints its low-metallicity signature on the new stars. In the present work, this behavior is consistently recovered across almost all scenarios. Except for Fig. \ref{fig:CGCG012_FSPS} (FSPS model for CGCG 012-070), all models and galaxies exhibit this central dilution, supporting the hypothesis of recent gas inflow. Taken together, the results provide evidence for a rejuvenation \citep{Riffel+09,Bessiere+17,Mallmann+18, Martin-Navarro+22, Riffel+23} of the stellar populations in AGN hosts, as indicated by the significant presence of young to intermediate-age populations in the nuclear regions of these sources, as recovered with the 3 different models.

\figsetstart
\figsetnum{5}
\figsettitle{Radial variations of the SP properties}
\figsetgrpstart
\figsetgrpnum{5.1}
\figsetgrptitle{Radial SP variations -- M13 models}
\figsetplot{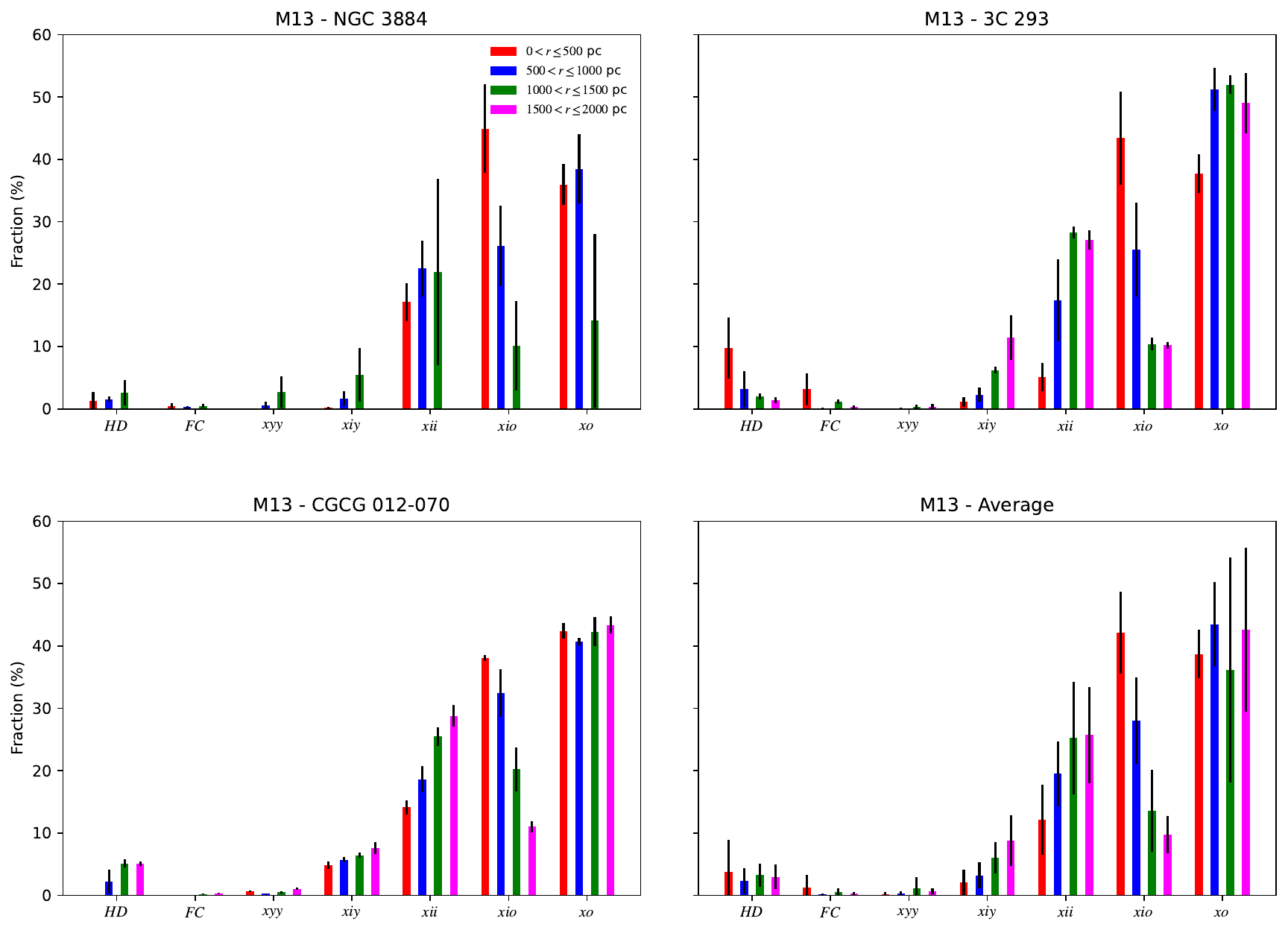}
\figsetgrpnote{Radial variations of the SP properties fitted with M13 models.}
\figsetgrpend
\figsetgrpstart
\figsetgrpnum{5.2}
\figsetgrptitle{Radial SP variations -- XSL models}
\figsetplot{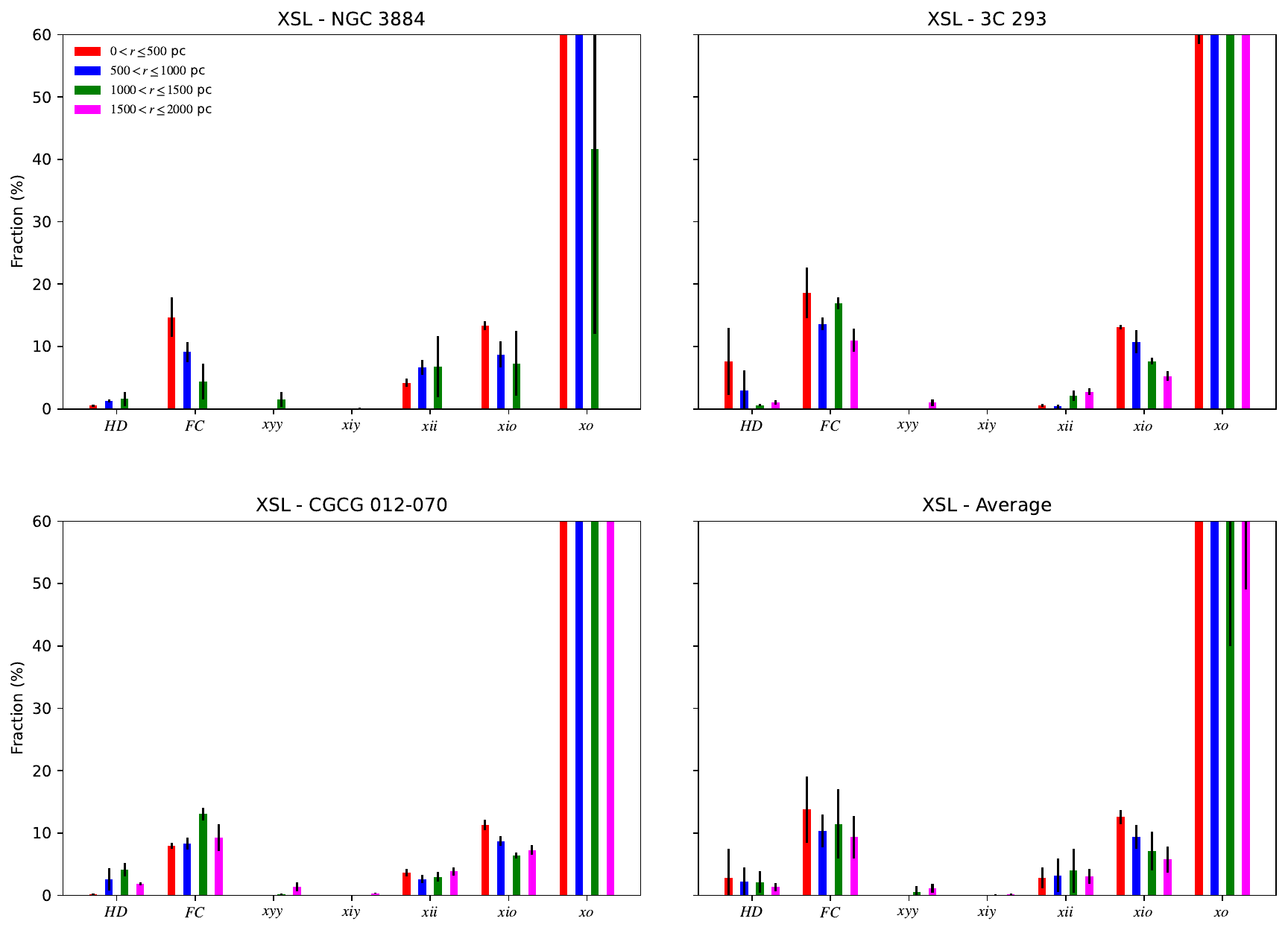}
\figsetgrpnote{Radial variations of the SP properties fitted with XSL models.}
\figsetgrpend
\figsetgrpstart
\figsetgrpnum{5.3}
\figsetgrptitle{Radial SP variations -- FSPS models}
\figsetplot{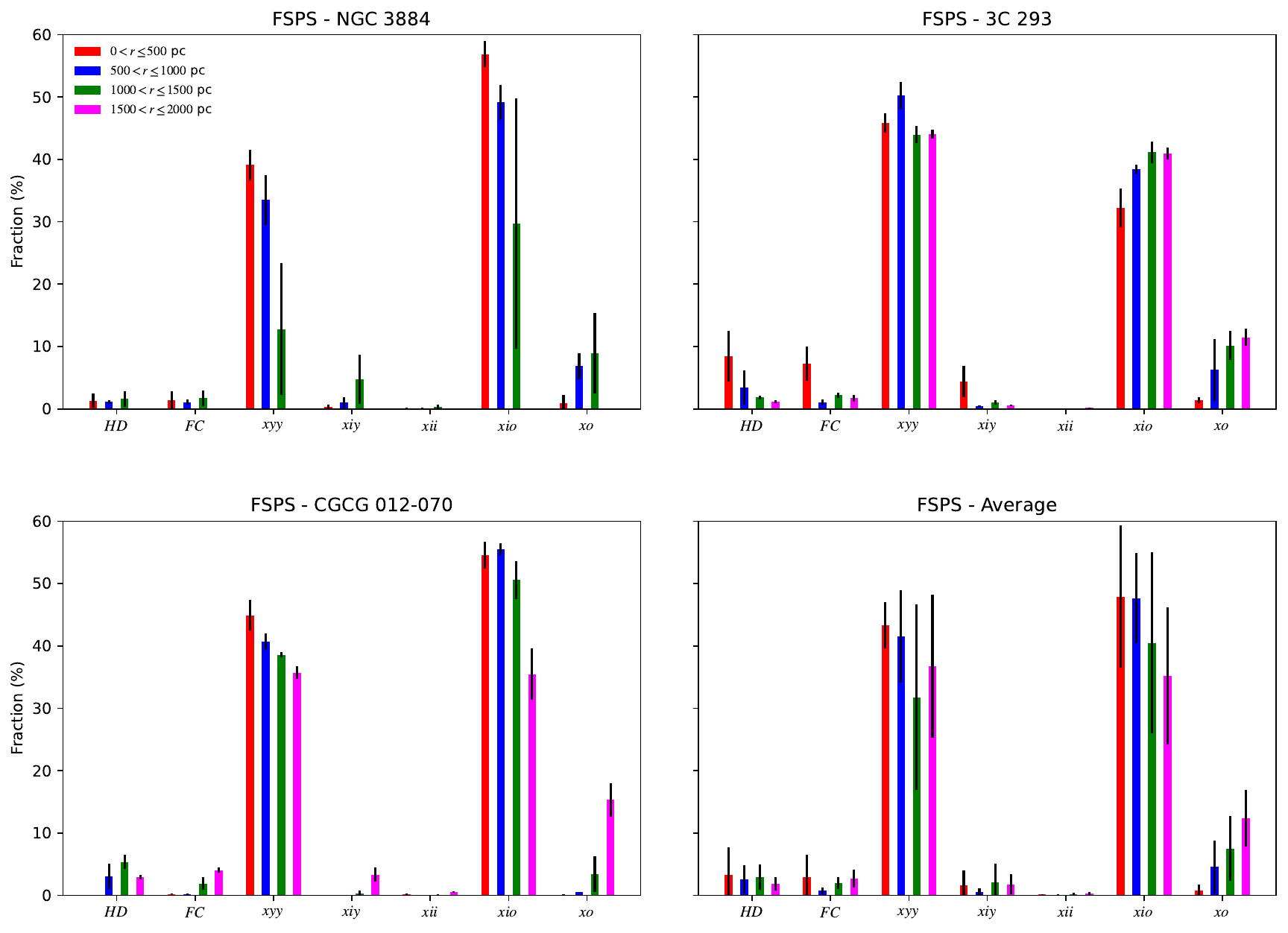}
\figsetgrpnote{Radial variations of the SP properties fitted with FSPS models.}
\figsetgrpend
\figsetend

\begin{figure*}
   \centering
   \includegraphics[width=1\linewidth]{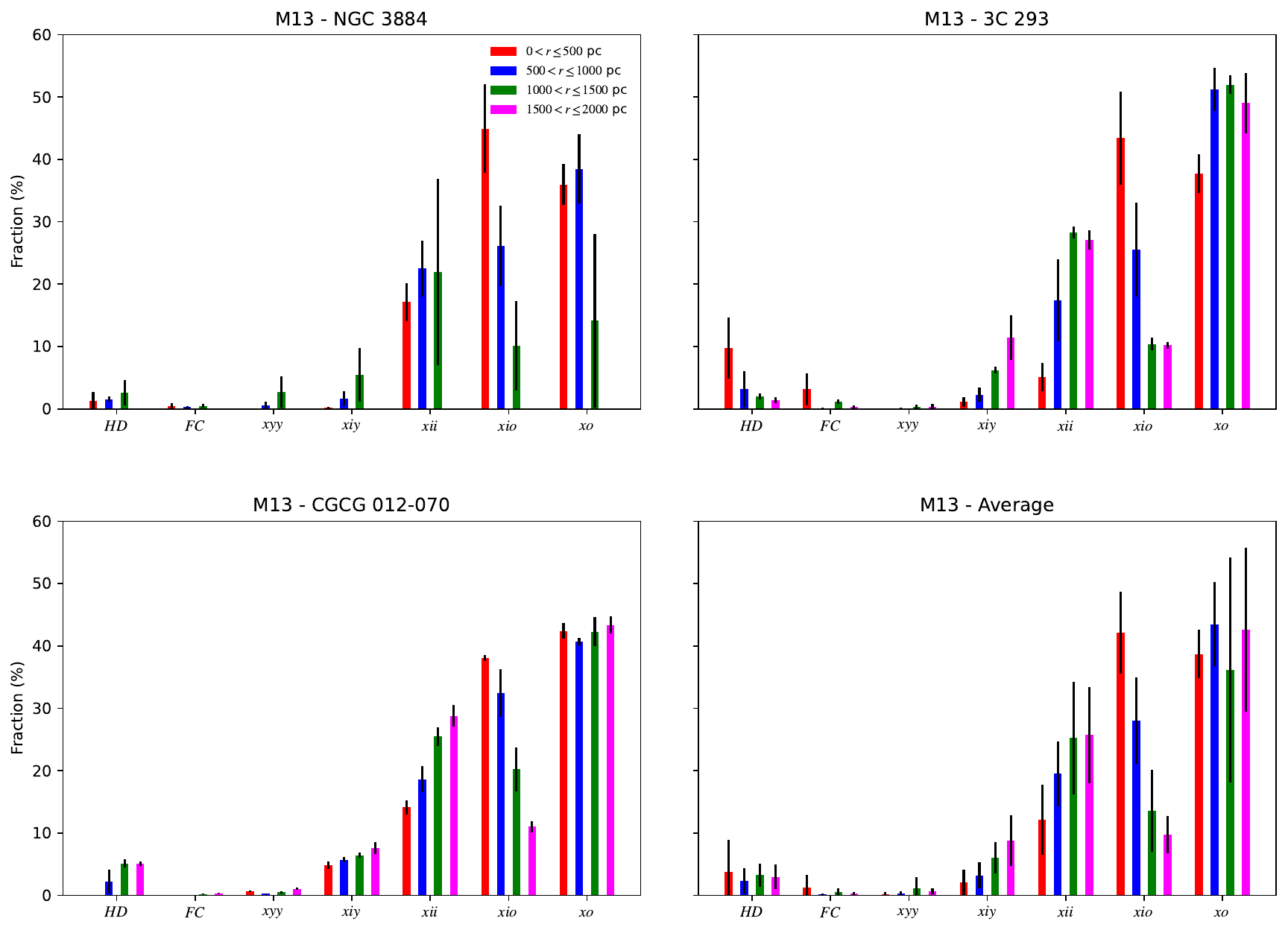}
       \caption{Radial variations of the SP properties fitted with M13 models. The complete figure set (3 images) is available in the online journal.}
   \label{fig:combinedM13}
   \label{fig:combinedXSL}
   \label{fig:combinedFSPS}
\end{figure*}

\citet{Gatto+25}, using the {\sc MEGACUBES}\footnote{\url{https://manga.if.ufrgs.br/}} from \citet{Riffel+23}, which are based on the MaNGA-SDSS datacubes \citep{Bundy+15} and employ the methodology developed in \citet{Riffel+21} and \citet{deMellos+24}, investigated the star formation rates (SFRs) in a sample of 293 AGNs and compared them with those of 492 carefully matched control galaxies (\citealp[for details on the control sample selection, see][]{Rembold+17, Riffel+23}). They found that, on average, AGN host galaxies exhibit central SFRs that are approximately a factor of two higher than those of their control counterparts. The most pronounced enhancements are observed in the most luminous AGNs and in hosts of the earliest morphological types, where the central SFR can reach values up to about four times those of the controls. 

Furthermore, they identified a strong correlation between the black hole accretion rate and the nuclear SFR, which supports the scenario in which both processes are fuelled by a common cold gas reservoir. When considering the global SFR integrated over the entire extent of each galaxy, they found that AGN hosts lie slightly below the star formation main sequence (SFMS). However, they did not interpret this offset as conclusive evidence for AGN-driven quenching, because the control galaxies lie even further below the SFMS. Instead, their results suggest that ongoing nuclear activity is more frequently associated with enhanced, rather than suppressed, star formation in the central regions of host galaxies.

In this framework, the presence of young and intermediate-age stellar populations implies that AGN fueling, or part of it, may occur with a time delay relative to the onset of star formation. As discussed by \citet{Riffel+22} and Marinho et al. (submitted), gas inflows toward the central regions can initially trigger circumnuclear star formation and/or AGN activity. As these stellar populations evolve, mass loss from massive stars and intermediate-age populations replenishes the nuclear gas reservoir, which can subsequently accrete onto the SMBH, enhancing or reactivating the AGN. Once the stellar populations age and the mass return diminishes, the gas supply to the nucleus is reduced, leading to a decline or shutdown of the AGN activity. A new cycle may then be initiated by fresh gas inflows. In fact, such recycled material has been invoked as a possible feeding channel in theoretical studies \citep{Choi+24} and observational ones \citep{Davies+07,Dahmer-Hahn+22,Riffel+11c,Riffel+22,Riffel+23,Riffel+24}.

\section{Conclusions}\label{sec:conclusions}

In this study, we have conducted a spatially resolved analysis of the stellar populations in the BAH sample of galaxies (NGC~3884, 3C~293, and CGCG~012-070) using data obtained with the JWST NIRSpec. By comparing the results derived from the M13, XSL, and FSPS stellar population models, we have constrained the star formation histories (SFHs) and the physical conditions of the nuclear and circumnuclear regions in these sources. Our main conclusions are as follows:

\begin{itemize}
    \item The stellar content of the galaxies in our sample is dominated by intermediate-age to old stellar populations with predominantly high metallicities ($Z \gtrsim 1\,Z_{\odot}$).
    
    \item We identify signatures of stellar population rejuvenation in these AGN host galaxies, as indicated by the substantial contribution of young to intermediate-age stellar components in the nuclear regions, consistently recovered by all three model sets.
    
    \item In NGC~3884 and 3C~293, the $xii$ component ($0.2 < t \leq 0.7$~Gyr) is frequently distributed in ring-like configurations surrounding the nucleus, suggesting a preferential circumnuclear locus for recent star formation.
    
    \item The unresolved nuclear emission exhibits substantial contributions from both a featureless continuum ($FC$) and hot dust ($HD$) components. In 3C~293, a significant $FC$ contribution is detected at two distinct spatial locations. While this configuration may be indicative of a dual active galactic nucleus (dual-AGN) system, an alternative explanation in which the secondary featureless component is produced by a heavily reddened starburst cannot be excluded.
    
    \item We detect a central decrease in stellar metallicity in almost all galaxies and models, supporting a scenario in which relatively metal-poor gas inflows have recently reached the nuclear regions, contributing to both AGN fueling and nuclear star formation.
    
    \item In agreement with previous works, we find that the $HD$ and $FC$ components decline with increasing galactocentric radius, while the relative contribution of younger stellar populations tends to increase toward the outer regions.
    
    \item Our results are consistent with a feeding–feedback cycle in which gas inflows trigger circumnuclear star formation, and the subsequent mass loss from evolving stars replenishes the nuclear gas reservoir, thereby sustaining or reactivating the AGN. This further supports models in which AGN fueling is at least partially regulated by recycled material from stellar evolution.
\end{itemize}

The overall agreement between the M13- and XSL-based results, both of which indicate the coexistence of evolved and relatively young stellar components, enhances the robustness of our conclusions. Nevertheless, the FSPS models systematically require a higher fractional contribution from very young stellar populations and (as in XSL library), yield more irregular, “bumpy” star formation history (SFH) solutions. In contrast, the M13-based analysis produces smoother and more continuous SFHs, as is typically expected for spiral galaxies. Collectively, these results highlight the pivotal importance of high–spatial–resolution near-infrared spectroscopy for disentangling the complex SFHs and nuclear structures of active galaxies.

%%%%%%%%%%%%%%%%%%%%%%%%%%%%%%%%%%%%%%%%%%%%%%%%%%

\begin{acknowledgments}
We thank the referee for the constructive comments, which improved the paper.
RR acknowledges support from  Conselho Nacional de Desenvolvimento Cient\'{i}fico e Tecnol\'ogico  ( CNPq, Proj. CNPq-445231/2024-6,311223/2020-6, 404238/2021-1, and 310413/2025-7), Funda\c{c}\~ao de amparo \`{a} pesquisa do Rio Grande do Sul (FAPERGS, Proj. 19/1750-2 and 24/2551-0001282-6) and Coordena\c{c}\~ao de Aperfei\c{c}oamento de Pessoal de N\'{i}vel Superior (CAPES, 88881.109987/2025-01).

RAR acknowledges the support from the Conselho Nacional de Desenvolvimento Cient\'ifico e Tecnol\'ogico (CNPq; Projects 303450/2022-3, and 403398/2023-1), the Coordena\c{c}\~ao de Aperfei\c{c}oamento de Pessoal de N\'ivel Superior (CAPES; Project 88887.894973/2023-00), and Funda\c{c}\~ao de Amparo \`a Pesquisa do Estado do Rio Grande do Sul (FAPERGS; Project 25/2551-0002765-9). 
MB acknowledges support from the Juan de La Cierva scholarship with reference JDC2023-052684-I, funded by MICIU/AEI/10.13039/501100011033 and FSE+. 
This work is based on observations made with the NASA/ESA/CSA James Webb Space Telescope. The data were obtained from the Mikulski Archive for Space Telescopes at the Space Telescope Science Institute, which is operated by the Association of Universities for Research in Astronomy, Inc., under NASA contract NAS 5-03127 for JWST. These observations are associated with program JWST-GO-01928. Support for program JWST-GO-01928 was provided by NASA through a grant from the Space Telescope Science Institute, which is operated by the Association of Universities for Research in Astronomy, Inc., under NASA contract NAS 5-03127. We acknowledge the utilization of artificial intelligence–based tools for the refinement and standardization of language.

\end{acknowledgments}

% \section*{Data Availability}
% The data will be made available under reasonable request. 

\bibliographystyle{aasjournal}
\bibliography{RiffelReferencesZotero}

\appendix

\section{Fits with other models}

These maps will just appear in the online version. 

\subsection{Results of XSL Model Fitting}

%The results obtained using the XSL models are presented in this subsection. Consistent with the trends observed in the previous sections, the XSL synthesis successfully reproduces the main spectral features of the NIR continuum across all three galaxies from the JWST NIRCam datacube and allows a direct comparison with the M13 and E-MILES solutions in terms of ages, metallicities, and non-stellar components.
\begin{figure*}
    \centering
    \includegraphics[width=0.85\linewidth]{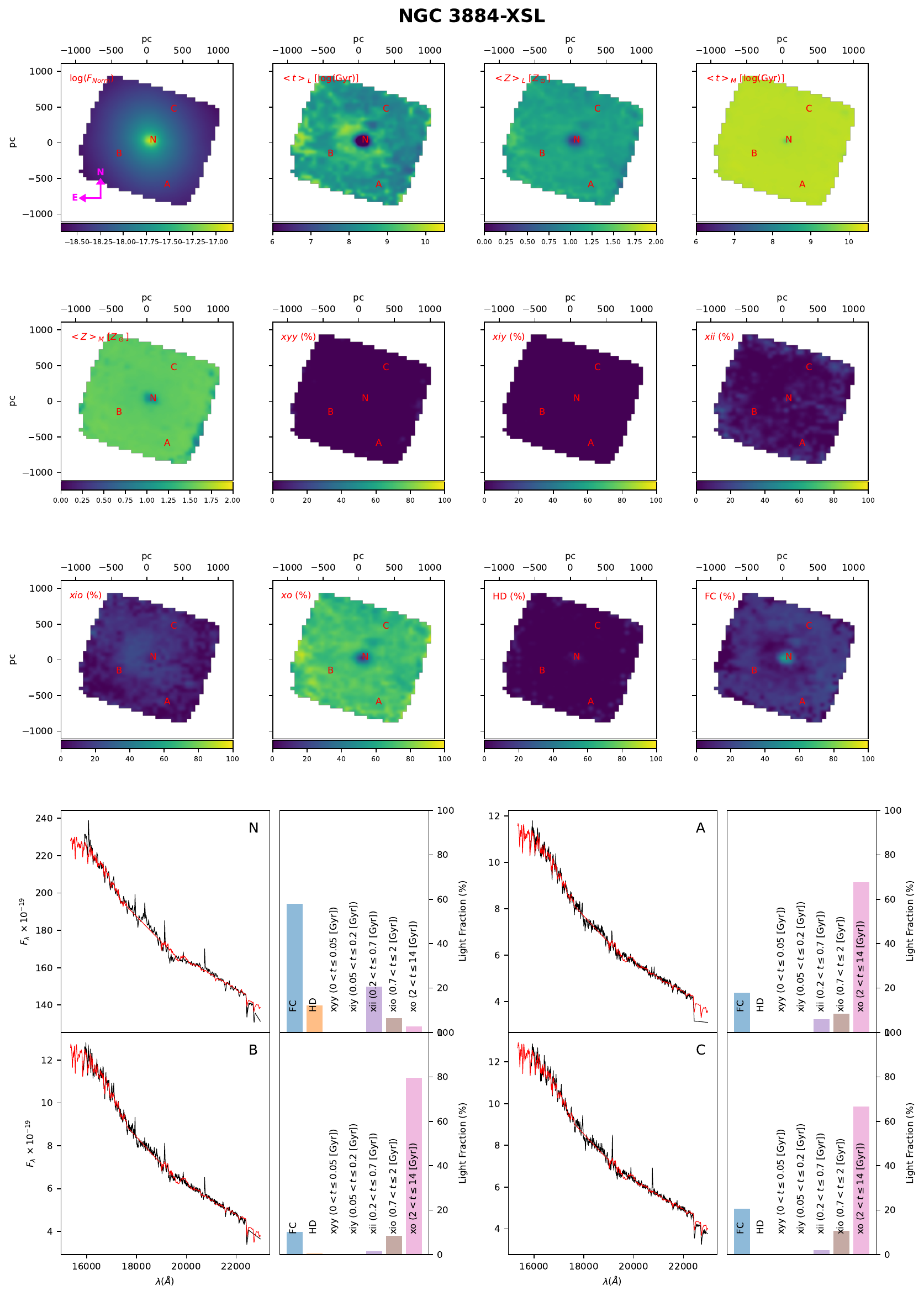} 
    \caption{Same as Fig.~\ref{3884M13plot} but for XSL models.}
    \label{fig:NGC3884_XSL}
\end{figure*}

\begin{figure*}
    \centering
    \includegraphics[width=0.85\linewidth]{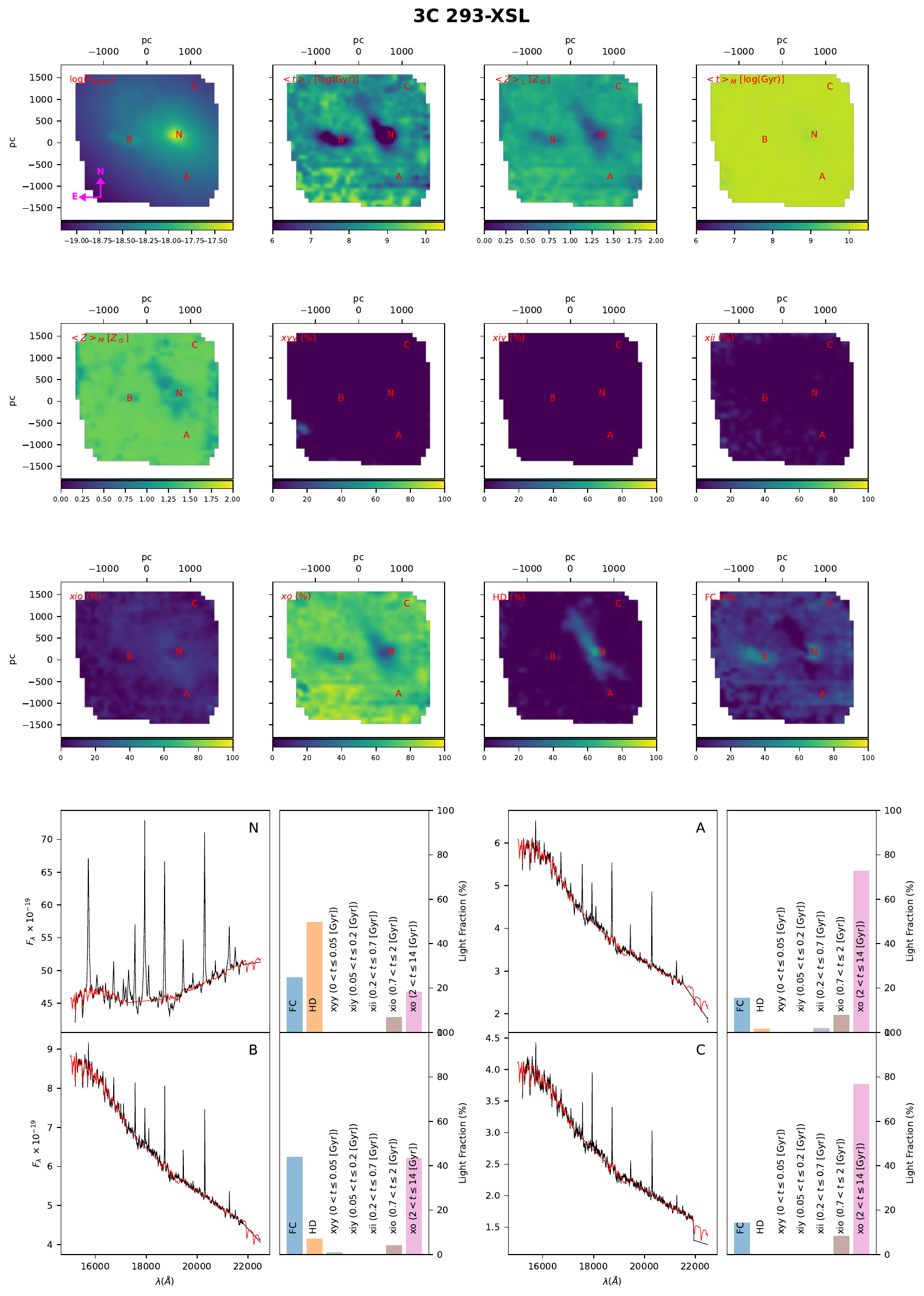} 
    \caption{Same as Fig.~\ref{3884M13plot} but for XSL models and for 3C~293.}
    \label{fig:3C_293_XSL}
\end{figure*}

\begin{figure*}
    \centering
    \includegraphics[width=0.85\linewidth]{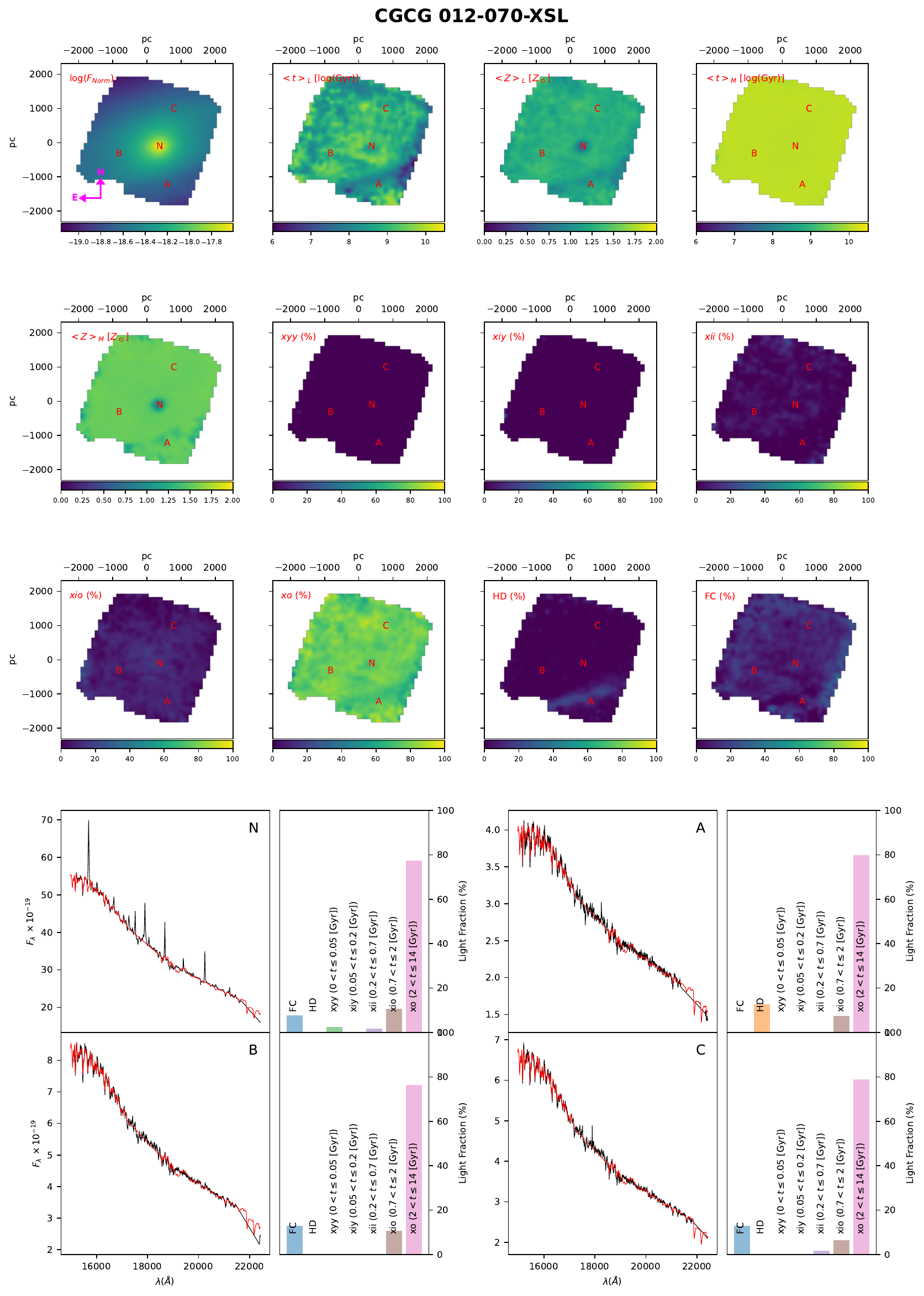} 
    \caption{Same as Fig.~\ref{3884M13plot} but for XSL models and CGCG 012-070.}
    \label{fig:CGCG012_XSL}
\end{figure*}

\subsection{Results of FSPS Model Fitting}

\begin{figure*}
    \centering
    \includegraphics[width=0.85\linewidth]{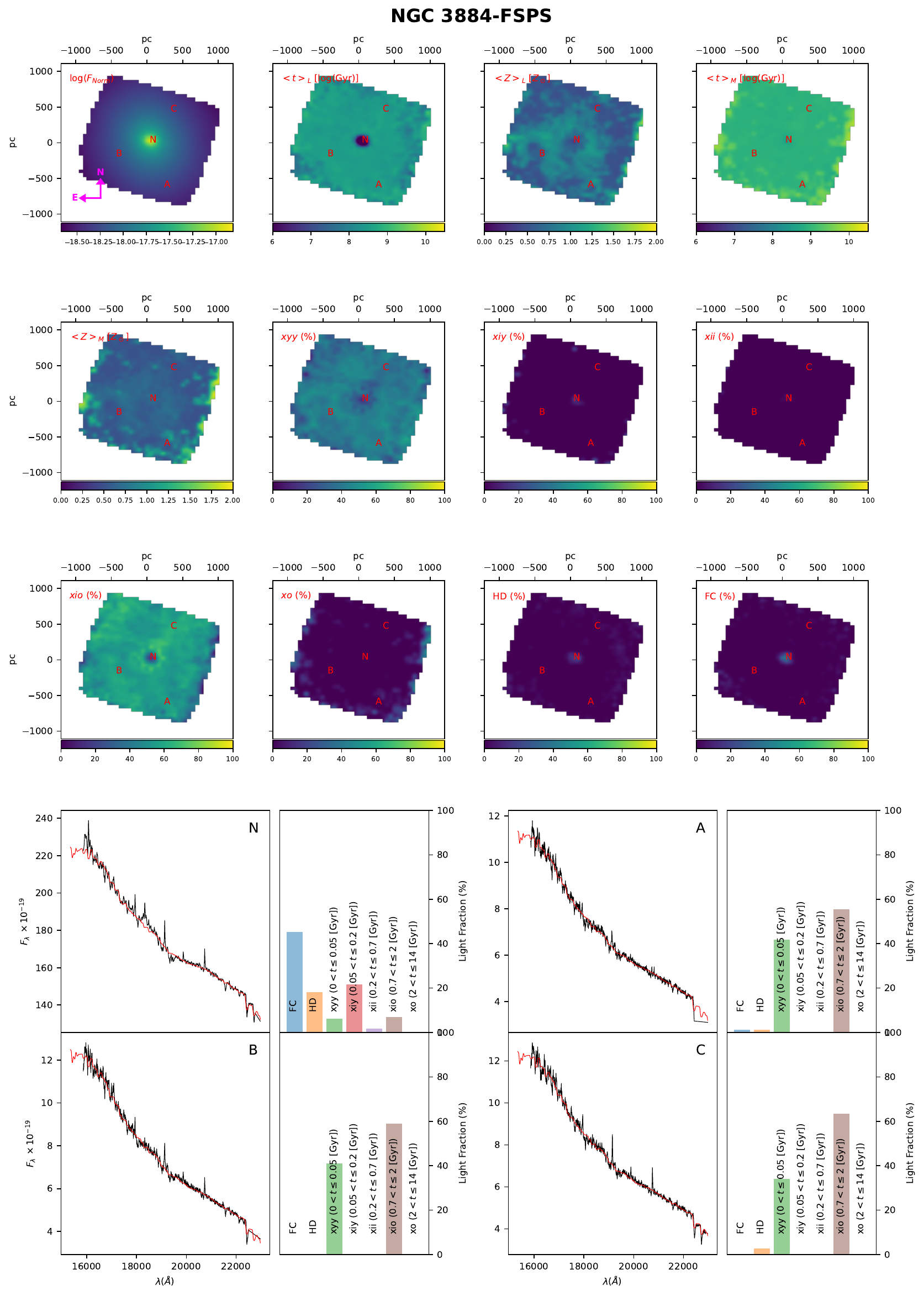}
    \caption{Same as Fig.~\ref{3884M13plot} but for FSPS models.}
    \label{fig:NGC3884_FSPS}
\end{figure*}

\begin{figure*}
    \centering
    \includegraphics[width=0.85\linewidth]{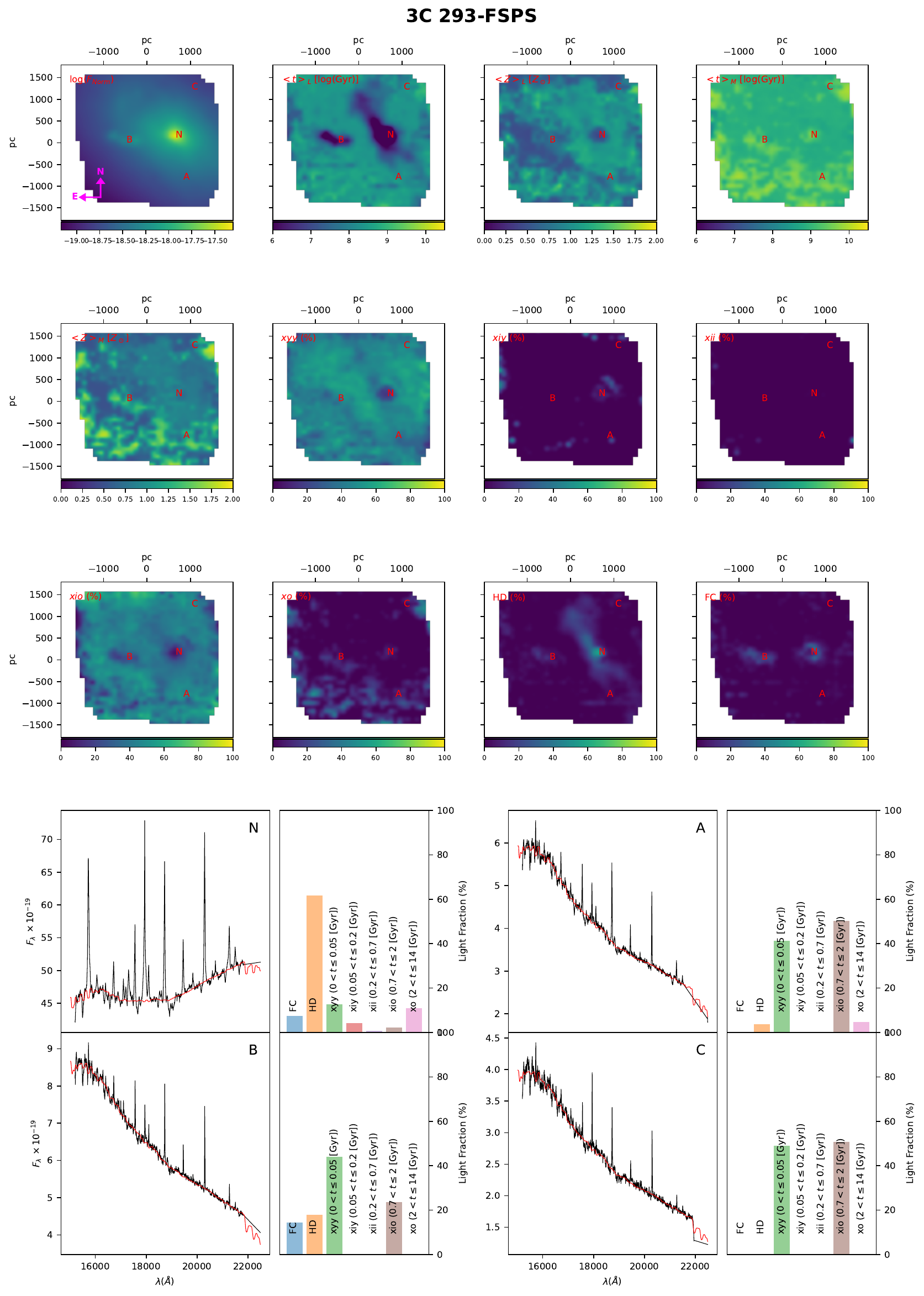}
\caption{Same as Fig.~\ref{3884M13plot} but for FSPS models and for 3C~293.}
    \label{fig:3C293_FSPS}
\end{figure*}

\begin{figure*}
    \centering
    \includegraphics[width=0.85\linewidth]{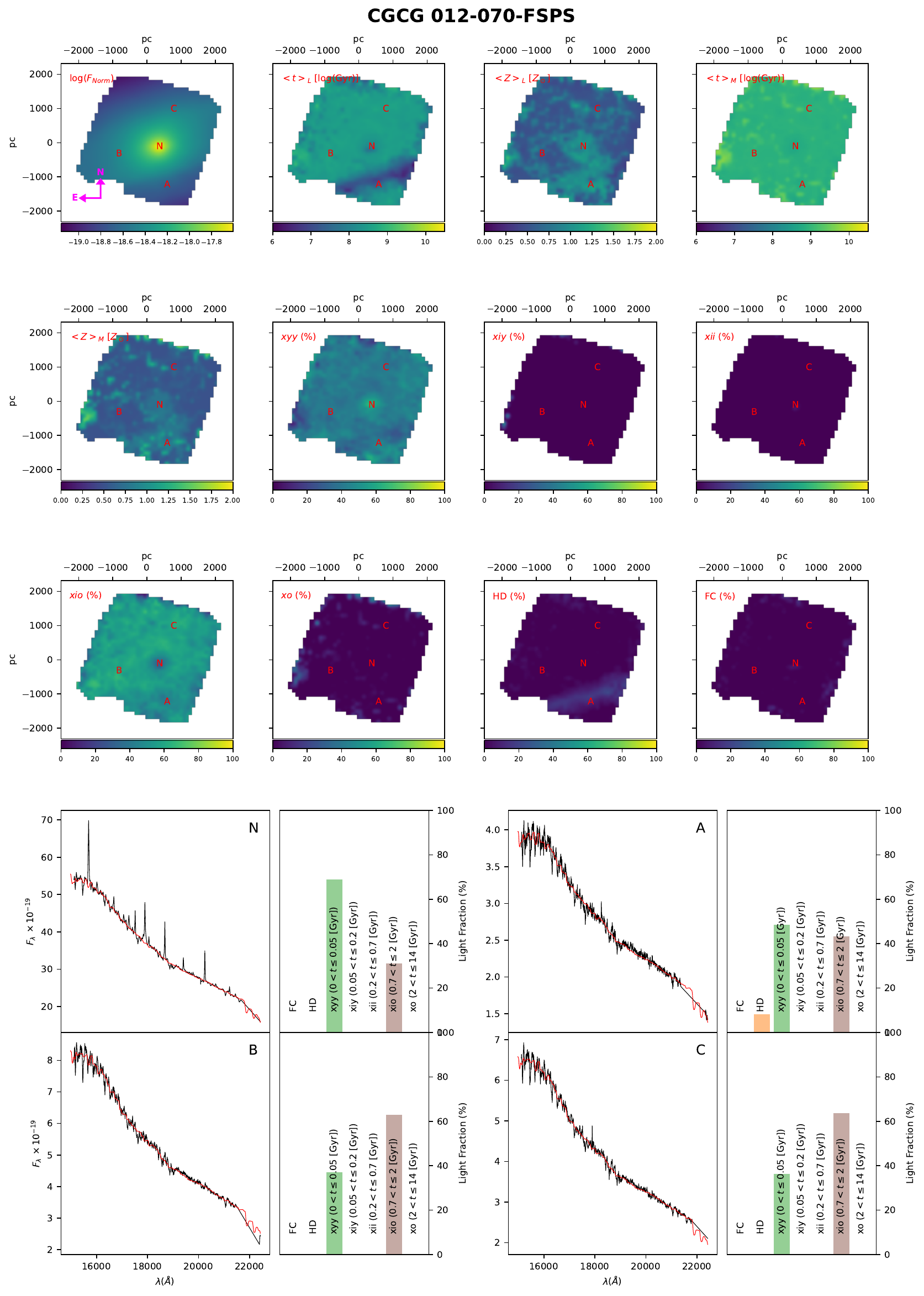}
\caption{Same as Fig.~\ref{3884M13plot} but for FSPS models and for CGCG 012-070.}
\label{fig:CGCG012_FSPS}
\end{figure*}

\begin{figure*}
   \centering
   \includegraphics[width=0.85\linewidth]{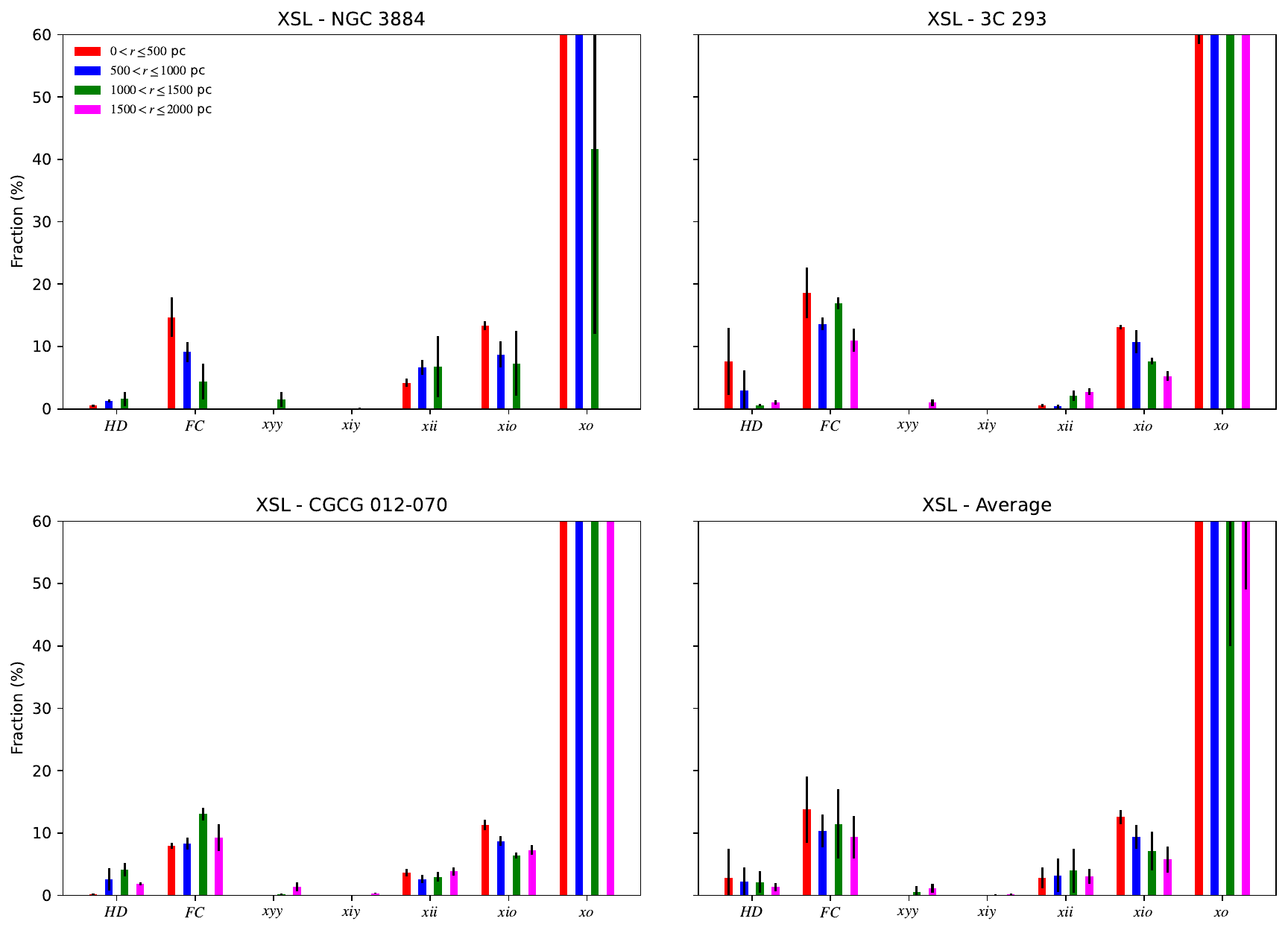} % Update path if necessary
       \caption{Radial variations of the SP properties fitted with XSL models.}
   \label{fig:combinedXSL}
\end{figure*}

\begin{figure*}
   \centering
   \includegraphics[width=0.85\linewidth]{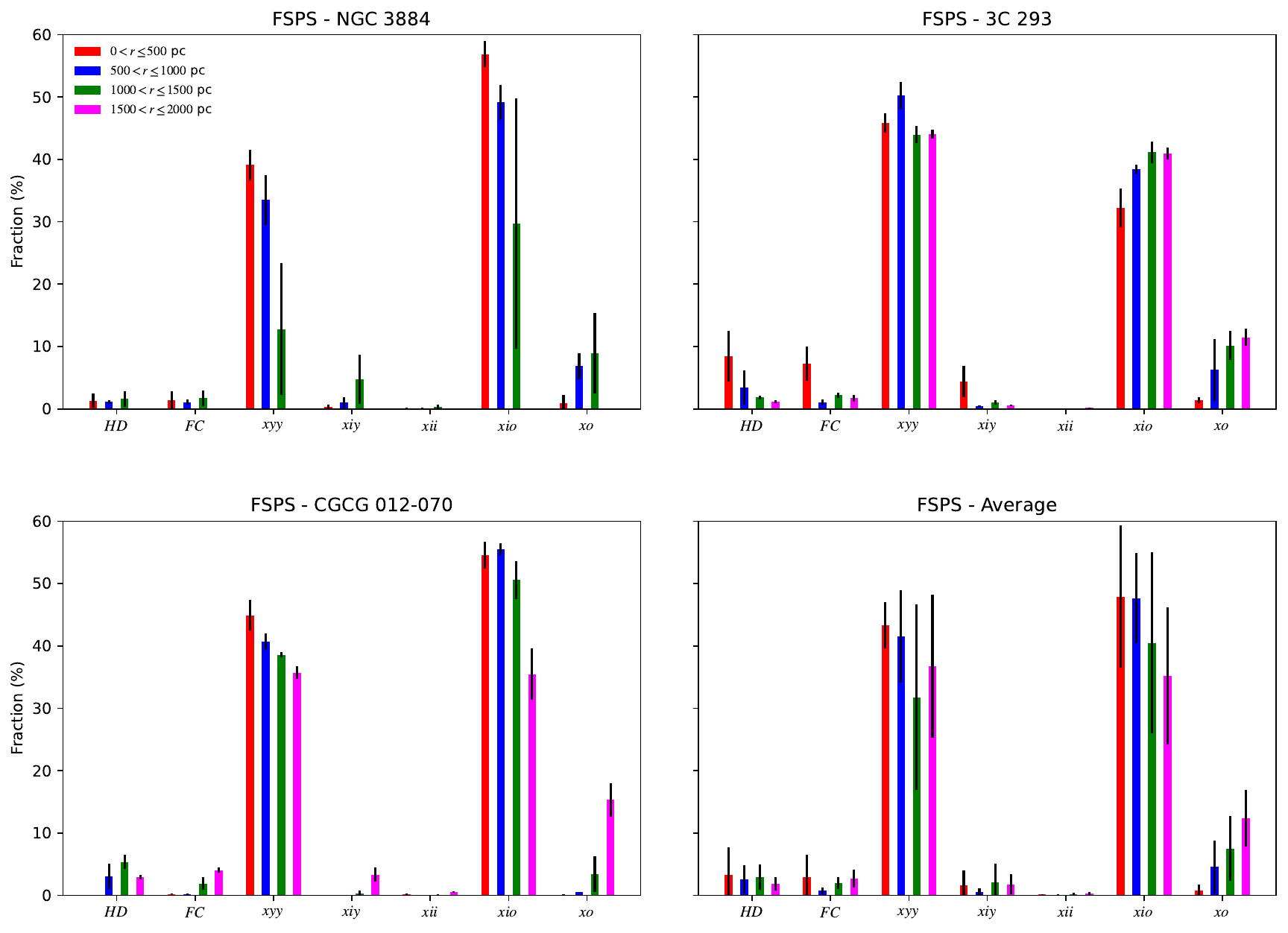} % Update path if necessary
       \caption{Radial variations of the SP properties fitted with FSPS models.}
   \label{fig:combinedFSPS}
\end{figure*}

\end{document}